\documentclass[useAMS,usenatbib]{mn2e}
\usepackage{times}
\usepackage{color}
\usepackage[fleqn]{amsmath}
\usepackage{aas_macros,amsfonts,amssymb}
\usepackage{url,hyperref}
\usepackage{graphicx}
\usepackage{booktabs}
\usepackage{siunitx}
\hypersetup{colorlinks=true,linkcolor=black,citecolor=black,filecolor=black,urlcolor=black}
\newcommand{\aCB}{\ensuremath{\alpha\,{\rm Cen}\,{\rm B }}}
\newcommand{\aCA}{\ensuremath{\alpha\,{\rm Cen}\,{\rm A }}}
\newcommand{\drv}{\ensuremath{\Delta {\rm RV}}}
\newcommand{\SHK}{\ensuremath{S _{\rm HK}}}
\newcommand{\lrhk}{\ensuremath{\log R'_{\rm HK}}}
\newcommand{\bis}{\ensuremath{{\rm BIS}}}
\newcommand{\bmg}[1]{\mbox{\boldmath $#1$}}
\newcommand{\bm}[1]{ \mathbf{#1} }
\newcommand{\cmps}{\ensuremath{\rmn{cm}~\rmn{s}^{-1}}}
\newcommand{\mps}{\ensuremath{\rmn{m}~\rmn{s}^{-1}}}
\newcommand{\kmps}{\ensuremath{\rmn{km}~\rmn{s}^{-1}}}
\title[A GP framework for modelling stellar activity]{A Gaussian process framework for modelling stellar activity signals in radial velocity data}
\author[V.~Rajpaul et al.]{V.~Rajpaul,$^{1}$\thanks{E-mail: vinesh.rajpaul@astro.ox.ac.uk} S.~Aigrain,$^{1}$ M.~A.~Osborne,$^{2}$ S.~Reece,$^{2}$ and S.~Roberts$^{2}$ 
\\
	$^{1}$Sub-department of Astrophysics, Department of Physics, University of Oxford, Oxford OX1 3RH, UK\\
 	$^{2}$Pattern Recognition and Machine Learning Group, Department of
 	Engineering Science, University of Oxford, Oxford OX1 3PJ, UK\\
 	}
\begin{document}
\date{Accepted 2015 June 24. Received 2015 June 22; in original form 2015 March 16}
\pagerange{\pageref{firstpage}--\pageref{lastpage}} \pubyear{2015}
\maketitle
\label{firstpage}
\begin{abstract}
To date, the radial velocity (RV) method has been one of the most productive techniques for detecting and confirming extrasolar planetary candidates. Unfortunately, stellar activity can induce RV variations which can drown out or even mimic planetary signals -- and it is notoriously difficult to model and thus mitigate the effects of these activity-induced nuisance signals. This is expected to be a major obstacle to using next-generation spectrographs to detect lower mass planets, planets with longer periods, and planets around more active stars. Enter Gaussian processes (GPs) which, we note, have a number of attractive features that make them very well suited to disentangling stellar activity signals from planetary signals. We present here a GP framework we developed to model RV time series jointly with ancillary activity indicators (e.g. bisector velocity spans, line widths, chromospheric activity indices), allowing the activity component of RV time series to be constrained and disentangled from e.g.\ planetary components. We discuss the mathematical details of our GP framework, and present results illustrating its encouraging performance on both synthetic and real RV datasets, including the publicly-available \mbox{Alpha Centauri B} dataset. 
\end{abstract}
\section{Introduction}\label{sec:intro}
The radial velocity (RV) method, a.k.a.\ Doppler spectroscopy, has been one of the most productive methods for discovering exoplanets. To date, more than 500 confirmed planets\footnote{Based on counts from the NASA Exoplanet Archive, available online at \url{exoplanetarchive.ipac.caltech.edu}.} have been discovered using the RV method, and until quite recently, more planets had been discovered with the RV method than with any other technique. Though more planets have now been discovered by transit photometry -- thanks to the unparalleled success of NASA's Kepler observatory in recent years \citep{batalha2013,marcy2014} -- the RV method remains indispensable both in its own right for discovering planets, and moreover for confirming and helping to characterise candidate planets detected by other means, including transit photometry.

Thanks to a number of technical developments, the precision of RV surveys has been constantly improving. Whereas the spectrographs of fifty years ago produced RV measurements with errors in excess of $1$~\kmps (\textcolor{black}{nominal precision per measurement}), today's cutting-edge, so-called ``second generation'' spectrographs can identify radial-velocity shifts down to a few tens of centimetres per second -- enough to locate many rocky, Earth-like planets orbiting close to their host stars \citep{pepe2011,lovis2011}. Examples of such spectrographs include the HARPS (High Accuracy Radial Velocity Planet Searcher) spectrograph hosted by the ESO $3.6$~m telescope in Chile; its Northern Hemisphere component, HARPS-N, on the Italian 3.58~m Telescopio Nazionale Galileo telescope on La Palma Island; and HIRES (High Resolution Echelle Spectrograph) at the Keck telescopes in Hawaii. Third generation spectrographs, to come online in the next few years, are expected to have measurement errors below $10$~\cmps \citep{pasquini2008,Pepe2010}.

This makes it possible, in principle, to discover exoplanets with increasingly low masses and increasingly longer periods, potentially reaching down to the Earth-mass and/or habitable regimes. This has motivated intensive, multi-season monitoring campaigns on current state-of-the-art instruments, and results from these campaigns are in turn bolstering the science cases for next-generation instruments to be deployed on the world's largest telescopes. 

One announcement in particular caused a significant stir in the exoplanet community: that of a planet around the star \aCB\, with an orbital period of $\sim 3.24$~days, and a minimum mass similar to that of the Earth, discovered using the HARPS instrument \citep{dumusque2012nature}. If confirmed, this discovery would be particularly remarkable not only because the minimum mass of the planet is so low (the lowest of any planet around a Sun-type star), but also because the host star is the nearest Sun-like star to our own system, and the RV semi-amplitude reportedly induced by the planet ($50$~\cmps) is the smallest to have been measured to date.

Unfortunately, though we live in an age of RV instruments with ever-improving precisions, better instrumentation has meant that exoplanet hunters are also being increasingly confronted by a significant obstacle: the contaminating presence of RV signals from stars themselves, which is arguably the most important source of noise (or rather nuisance signal) in modern RV datasets. 

This, then, establishes the broad context for the present work, the focus of which is a Gaussian process framework we developed for modelling the activity-induced variations in RV time series, in order to disentangle activity from e.g.\ planetary contributions.

The balance of our paper is structured as follows. In the next section (Section \ref{sec:nuisance}), we discuss these stellar nuisance signals in some more detail; we note that signals related to stellar activity are particularly difficult to mitigate. In Section \ref{sec:formalism} we present our new framework for modelling stellar activity signals jointly with possible planetary signals, and discuss the potential advantages and shortcomings of our new framework. In Section \ref{sec:tests}, we demonstrate the successful application of our new framework to both simulated and real datasets, including the publicly available \aCB\ dataset. Finally, we present our conclusions in Section \ref{sec:discuss}, and discuss future plans for extending this work.
\section{Stellar nuisance signals}\label{sec:nuisance}
 In order of increasing associated time-scales, the main physical phenomena in stars giving rise to RV perturbations are the following (see e.g.\ \citealt{dumusque2012} for a detailed overview).
\begin{enumerate}
\item Oscillation of the external envelopes in Sun-like stars. Interference of multiple \mbox{$p$-mode} oscillations can introduce RV variations on the order of tens of centimetres per second, over time-scales of minutes \citep{schirjver2000,otoole2008,michel2008}.
\item Granulation phenomena, driven by convective flows in the external layers of Sun-like stars. Time-scales can range from several minutes to up to two days (in the case of super-granulation), with associated disk-integrated RV signals on the order of meters per second \citep{dravins1987,michel2008,mathur2011}.
\item Rotationally-modulated phenomena associated with active regions -- dark spots and bright plages and faculae -- which move across the observed stellar disk and break the flux balance between the redshifted and blueshifted halves of the disk \citep{dumusque2011b,boisse2012}. The major contribution to RV perturbations arises not directly from the presence of the spots and plages themselves, but from the strong magnetic fields associated with these active regions, which in turn lead to a suppression of the net blueshift normally associated with convective cells \citep{meunier2010}. The strength of associated RV perturbations depends on rotation and surface activity levels, and can be as high as tens of meters per second \citep{saar1997}; observations suggest that even chromospherically-quiet stars can be expected to have activity-induced RV signals on the order of \mbox{1--2~\mps} \citep{isaacson2010}. The associated time-scales are linked to the rotation period of the star, which can range from several hours to days or even weeks, as is the case for the Sun \citep{tassoul2000,nielsen2013}.
\item Long-term magnetic activity cycles. In Sun-like stars with long-term magnetic cycles, long-term modulations of activity levels and numbers of surface magnetic regions can lead to variations in stellar-origin RV perturbations of up to tens of meters per second over the course of many years \citep{dasilva2012,meunier2013}.
\end{enumerate} 

While observational strategies can be devised to mitigate nuisance RV signals associated with stellar oscillation and granulation \citep{dumusque2011a}, activity-induced signals pose a bigger challenge. Apart from the larger amplitude associated with these signals, their periodic or quasi-periodic nature means they can sometimes \emph{mimic} planetary RV signals (e.g.\ \citealt{desidera2004}, \citealt{bonfils2007}; \citealt{carolo2014}; \citealt{haywood2014b}; \citealt{santos2014}). Moreover, the associated time-scales of these stellar signals can be comparable to those expected for planets: on the order of days or weeks for rotationally-modulated activity, and on the order of years for periodic modulations due to long-term activity cycles. Therefore, in the regime of small-amplitude RV variations associated with lower-mass planets, planetary detection and characterisation rapidly becomes limited not by the quality of available spectrographs, but rather by nuisance signals which can be very difficult to disentangle from putative planetary signals. The Earth, for example, induces a reflex motion in the Sun of about $0.09$~\mps; however, though such a small RV signal could in principle be detected with next-generation spectrographs, in practice it would be masked by larger, activity-induced signals from the sun \citep{meunier2010,dumusque2012}.

There has thus been considerable interest, in recent years, in developing ways to mitigate the impact of activity, using a variety of methods ranging from exploiting simple correlations between the radial velocity measurements and chromospheric activity or line shape indicators \citep{boi+09,tuomi2014,robertson2014}, pre-whitening of the RV time series \citep{que+09}, and using red-noise models \citep{tuomi2013,feroz2014}, to fitting sine waves at the rotational period of the star and its harmonics \citep{boisse2011,dumusque2012nature} and detailed modelling of stellar surface features \citep{lan+10,boisse2012}. Each of these approaches, though, has drawbacks. For example, since activity variations cannot be expected to be strictly periodic, modelling them using sinusoids (which are not orthonormal in the case of unevenly sampled data) based on a na\"ive periodogram analysis can introduce harmonics of the subtracted sinusoids into the data, and bias one way from genuine signals in the data \citep{tuomi2014}.

\citet[hereafter A12]{aigrain2012} showed that there should exist a relatively simple relationship between the photometric brightness and radial velocity variations of a spotted star. This leads to a straight-forward method to predict the RV variations from the light curve, with only two free parameters. Because it uses the stellar flux and its time derivative, this method was named the $FF$' method, but it can be applied only when simultaneous high-precision photometric and RV measurements are available.

When this is not the case, information can still be gained by modelling the RV time series jointly with other parameters, which are extracted simultaneously from the stellar spectra. In particular, the width (FWHM) and degree of asymmetry of the spectral lines (usually measured from the CCF bisector span or bisector inverse slope), as well as chromospheric activity indicators (e.g.\ \lrhk, which measures the amount of emission in the cores of the Calcium {\sc ii} H \& K lines), are all sensitive to activity, but not to the presence of planets. \citet{pon+11} were among the first to carry out explicit joint modelling of RVs and these ancillary activity indicators, in the context of CoRoT-7. They fit a simple spot model to the FWHM and bisector span time series and used it to predict the RV variations due to activity. However, spot models are always very degenerate, making a robust exploration of the parameter space of the model impractical.

In the next section we introduce an extension of the $FF$' method, which uses the same principles, but models the RVs jointly with one or more of the aforementioned activity indicators. The various time series are modelled as linear combinations of a single, underlying Gaussian process (GP) and its time derivative, plus a separate noise term which is specific to each time series. 
\section{Proposed formalism}\label{sec:formalism}
\subsection{Formulation of the physical model}\label{sec:physical-model}
The physical model we propose to use is an extension of the $FF'$ framework introduced in A12, which can be used to estimate RV variations due to stellar activity using photometry. In this framework, the fraction of the visible stellar hemisphere that is covered in spots is represented by $F(t)$. For the simplest case of a single, small, equatorial spot, $F(t)$ represents the projected area of the spot, and varies as $\cos( 2 \pi t / P)$, where $P$ is the rotation period. A12 then showed that the RV perturbation due to activity (\textcolor{black}{even for a non-equatorial spot}) can be written: 
\begin{equation} 
\label{eq:FFprime1}
\drv = V_{\rm r} \, F(t) \, \dot{F}(t) + V_{\rm c} \, F^2(t),
\end{equation} 
where the first term represents the rotational modulation signal, and the second term the suppression of convective blueshift in magnetized regions. This remarkably simple expression arises because the radial velocity of the stellar surface varies as $\sin(2 \pi t / P)$, which is proportional to $\dot{F}(t)$, whereas the projection of the convective upwelling along the line of sight varies as $\cos(2 \pi t / P)$, which is proportional to $F(t)$. Both terms are multiplied by the projected spot area, which is proportional to $F(t)$. The two constants $V_{\rm r}$ and $V_{\rm c}$ can be related to physical quantities such as the fractional coverage and contrast of spots and faculae, the stellar equatorial velocity and the convective blueshift velocity, but are treated as free parameters for the present purpose. Equation~\eqref{eq:FFprime1} was derived by considering a single active region, but is used to describe the combined effect of all the active regions on the visible hemisphere. Though this is a first-order approximation, A12 carried out a number of tests of this formalism and showed that its performance was similar to that of more sophisticated methods. A caveat, however, is that photometry is insensitive to certain spot distributions, i.e.\ photometric signals will not always contain sufficient information adequately to predict RV variations. This problem is addressed in this paper, where we propose to use more than one activity indicator to constrain the activity component in RV variations.

We thus seek similar expressions to estimate RV variations due to stellar activity using activity diagnostics \emph{other than photometry}. We consider the case where we have a chromospheric activity indicator and a line asymmetry measurements available; for the sake of concreteness, we take these to be the \lrhk~index and the inverse slope of the bisector of the cross-correlation function, or simply bisector inverse slope (\bis), respectively, as is the case for HARPS datasets. 

As a first approximation, it seems reasonable to expect \lrhk\ and to behave as $F^2(t)$, the convective blueshift suppression term, which is close proxy for the fractional coverage of active regions. On the other hand, like the RVs, the bisector inverse slope  also depends on the velocity of the stellar surface at the location of active regions, so the expression for the BIS should include an additional $F(t) \, \dot{F}(t)$ term: 
\begin{eqnarray} 
\label{eq:lrhk} \lrhk & = & L_{\rm c} \, F^2(t) \\ 
\label{eq:bis} \bis & = & B_{\rm r} \, F(t) \, \dot{F}(t) + B_{\rm c} \, F^2(t), 
\end{eqnarray} 
where $L_{\rm c}$, $B_{\rm c}$ and $B_{\rm c}$ are free parameters. Additionally, each time series will have its own noise term, which is treated as white (this is discussed in more detail in later sections).

Equations \ref{eq:lrhk} and \ref{eq:bis} relate two particular activity-sensitive observables (\lrhk\ and \bis) to spot coverage, although analogous relations could be constructed for other activity-sensitive observables. For example, the \SHK\ chromospheric activity index can be related to spot coverage in the same way as the \lrhk\ index \citep{isaacson2010}. Similarly, FWHM data are usually very tightly correlated with (if noisier than) \lrhk\ data -- especially if there is a tight relation between the size and location of the photospheric spots and that of the chromospheric structures such as faculae -- and so the same form of relation could be used.
\subsection{Gaussian process framework}
Gaussian processes provide a mathematically-tractable and very flexible framework for performing Bayesian inference about functions. They are particularly suitable for the joint modelling of deterministic processes with stochastic processes of unknown functional forms (though with some known properties); they lend themselves very naturally to rigorous error propagation; and they allow us to relax the usually quite restrictive (e.g.\ linear) relationships which have been assumed in previous attempts at the joint modelling of RV with ancillary times-series \citep{tuomi2014}.

GP regression enables us to model complex stochastic processes by parametrising the covariance between pairs of datapoints, rather than writing down an expression for the data themselves; deterministic components of the model can also be incorporated as the mean of the GP. Mathematically, the prior joint distribution of the outputs $\bm{y}$ (observables, in the case of our framework) is taken to be a multi-variate Gaussian:
\begin{equation}
\label{eq:GP}
 p(\bm{y}) = \mathcal{N}( \bm{y}; \bm{m}, \bm{K}).
\end{equation} 
The mean vector $\bm{m}$ is given by
\begin{equation}
\label{eq:mean}
 \bm{m} = m(\bm{x};\bmg{\theta}),
\end{equation} 
where $m$ is a mean function of the inputs $\bm{x}$ (observation times, for our framework) with parameters $\bmg{\theta}$. The elements of the covariance matrix $\bm{K}$ are given by
\begin{equation}
\label{eq:kernel}
 \bm{K}_{ij} = k(x_i, x_j;\bmg{\phi}),
\end{equation} 
where $k$ is a covariance kernel function with parameters $\bmg{\phi}$. The covariance kernel function $k$ provides the covariance element between any two sample locations or times, $x_i$ and $x_j$.

The elements of $\bm{m}$ and $\bm{K}$ are, in the strict sense of the term, the parameters of the model; however, the values of these parameters are controlled by a small number of \emph{hyper-parameters} $\bmg{\theta}$ and $\bmg{\phi}$, and in practice will never be inferred directly. 

Gaussian distributions obey many convenient mathematical identities, which allow us trivially to marginalise over unobserved function values -- even in the common case where there are an infinite number of such values. This remarkable property enables us to write down a marginal likelihood $\mathcal{L}_{m,k}(\bmg{\theta},\bmg{\phi})$ for the data, given a GP model: 
\begin{equation}
\log \left[ {{\mathcal{L}_{m,k}}(\theta ,\phi )} \right] =  - \frac{1}{2}{{\mathbf{r}}^{\text{T}}}{{\mathbf{K}}^{ - 1}}{{\mathbf{r}}^{\text{T}}} - \frac{1}{2}\log \left( {\det {\mathbf{K}}} \right) - \frac{n}{2}\log \left( {2\pi } \right),
\end{equation} 
where $\mathbf{r}= {\mathbf{y}} - {\mathbf{m}}$ is the vector of residuals of the data after the mean function has been subtracted, and $n$ is the number of datapoints. This process is known as GP regression, and is described in more detail in textbooks such as \citet{bishop2006} and \citet{rasmussen2006}; alternatively, see \citet{roberts2013} for an accessible, tutorial-style introduction to GP regression. The free \emph{hyper-parameters} $\bmg{\theta}$ and $\bmg{\phi}$ can then be varied to maximise $\mathcal{L}_{m,k}$; this process is known as Type-II maximum likelihood, or marginal likelihood maximisation \citep{gibson2012}. In so doing we refine vague distributions over many, very different functions -- the forms of which are controlled by $\bmg{\theta}$ and $\bmg{\phi}$ -- to more precise distributions which are focused on functions that best explain our observed data. 

To compare models with different mean and covariance functions ($m$ and $k$, respectively), we must compute the model evidence for each model, which involves numerical integration of the likelihood $\mathcal{L}_{m,k}$ and priors over hyper-parameters $\bmg{\theta}$ and $\bmg{\phi}$. In practice, the computational cost of doing this can be extremely expensive. Furthermore, formulating physically motivated, statistically proper priors for each hyper-parameter can be challenging. On the other hand, \textcolor{black}{it is often both theoretically-defensible and computationally-convenient \citep[see][]{gibson2012} first to maximise $\mathcal{L}_{m,k}$ with respect to all hyper-parameters, next to fix the `uninteresting' ones (typically those in $\bmg{\phi}$, which are often also degenerate) to their \textcolor{black}{maximum \emph{a posteriori} (MAP)} values, and lastly to marginalise fully over the remaining parameters of interest (typically those in $\bmg{\theta}$). This approximation avoids having to perform matrix inversion at every step of marginalisation -- in practice, this entails marginalising over planet parameters, for example, but simply optimising the parameters of covariance kernels. We adopt this approximation throughout this paper.}
\subsection{GP model for multiple time series}\label{sec:GP-model}
Again for the sake of concreteness, we consider the case of a HARPS-like dataset, where we might wish to model jointly three observables: RV perturbations \drv, an activity index \lrhk, and the bisector inverse slope \bis. Further, to illustrate the way deterministic components can be incorporated into the framework, we assume -- as is the case for the \aCB\ dataset, analysed in Section \ref{sec:tests-alphaCen} -- that we wish to model dynamical effects in the RVs (binary motion, possible planet, etc.). 

In this case, we have $N$ observations of one independent variable, time $t$, and three dependent variables: \drv, \lrhk, and \bis. In order to treat all three as different manifestations of a single underlying process, we re-arrange the data into $3N$ observations of a {2-dimensional input $\bm{x} = [t, l]$}, and of the corresponding output $y$, where $y^{(l)}(t) = \drv$ if $l=1$, \lrhk\ if $l=2$ and \bis\ if $l=3$.

The baseline mean function is taken to be a second order polynomial, which in this example would suffice to represent the motion of \aCB\ around the centre of mass of the $\alpha\,$Cen binary system \citep{dumusque2012nature}, for the RV data, and a constant (DC offset) for the other time series:
\begin{eqnarray}
m(\bm{x}) & = & m^{(l)}(t),\\
m^{(1)}(t) & = & a + bt + ct^2, \\
m^{(2)}(t) & = & d, \\
m^{(3)}(t) & = & e, \\
\end{eqnarray}
where $\bmg{\theta} = [a, b, c, d, e]$ are free hyper-parameters. One or more Keplerian terms (each with an additional five hyper-parameters) can be added to the mean function to represent planetary signals, but this baseline model has no planet.

A white-noise component (to encapsulate observational noise, \textcolor{black}{and also activity-induced and other stellar signals not captured by our model, arising e.g.\ from pulsation}) is incorporated in the covariance function as follows:
\begin{equation}\label{eq:whitenoise}
{k(\bm{x}_i,\bm{x}_j)} = k_{\rm act}^{(l_i,l_j)}(t_i,t_j) + \sigma^2_{l_i} \, \delta(t_i-t_j)
\end{equation}
where $\sigma_{l_i}$ is the standard deviation of the noise associated with each type of observation, and $\delta(x)$ is the Kronecker delta function.

We next introduce a new latent (unobserved) variable, $G(t) \equiv F^2(t)$, which represents the activity term, and is described by a zero-mean Gaussian process with covariance function ${\gamma}$. Conveniently, we can then re-write equations~\eqref{eq:FFprime1}, \eqref{eq:lrhk}, and \eqref{eq:bis} as
\begin{eqnarray}
\drv & = & V_{\rm c} \, G(t) + V_{\rm r} \, {\dot{G}(t)}, \label{eq:phys-model-1} \\
\lrhk & = & L_{\rm c} \, G(t), \label{eq:phys-model-2} \\
\bis & = & B_{\rm c} \, G(t) + B_{\rm r} \, {\dot{G}(t)}, \label{eq:phys-model-3}
\end{eqnarray}
where a factor 2 has been incorporated into the constants $V_{\rm r}$ and $B_{\rm r}$. We now make use of the fact that any linear combination of a GP and its derivative\footnote{In fact, \emph{any} affine operator (including linear, derivative, and integral operators) applied to a GP yields another GP. This enables our framework to exploit the well-known analytic identities for conditioning and marginalising Gaussian distributions, and makes the problem computationally tractable.} is also a GP; following \citet{osb10}, the covariance between an observation of $G$ at time $t_i$ and an observation of {$\dot{G}$} at time $t_j$ is
\begin{equation}
{{\gamma ^{(G,dG)}}({t_i},{t_j}) = \frac{\partial }{{\partial t}}{\left. {{\gamma ^{(G,G)}}(t,{t_j})} \right|_{t = {t_i}}},}
\end{equation}
{where ${{\gamma ^{(G,G)}}({t_i},{t_j})}$ is used to denote the covariance between (non-derivative) observations of $G$ at times $t_i$ and $t_j$. Similarly, the} covariance between two observations of {$\dot{G}$} at times {$t_i$ and $t_j$} is
\begin{equation}
{{\gamma ^{(dG,dG)}}({t_i},{t_j}) = \frac{\partial }{{\partial t'}}\frac{\partial }{{\partial t}}{\left. {{{\left. {{\gamma ^{(G,G)}}(t,t')} \right|}_{t = {t_i}}}} \right|_{t' = {t_j}}}}.
\end{equation}

{Using the above notation, the individual covariance functions over the various inputs can then be related to $\gamma$} as follows (in the notation below, $k_{\rm act}^{(12)}=\operatorname{cov} \left( {\Delta {\text{RV}},\lrhk}\right)$, $k_{\rm act}^{(31)}=\operatorname{cov} \left( {\bis},\Delta \text{RV}\right)$, etc.):
{
\begin{eqnarray}
k_{\rm act}^{(11)}(t_i,t_j) & = & V_{\rm c}^2 \, {\gamma ^{(G,G)}}({t_i},{t_j}) + V_{\rm
  r}^2 \, {\gamma ^{(dG,dG)}}({t_i},{t_j}) \nonumber \\ &&+ \, {V_{\rm{c}}}\,{V_{\rm{r}}}\left( {\,{\gamma ^{(G,dG)}}({t_i},{t_j})
   + {\gamma ^{(dG,G)}}({t_i},{t_j})} \right), \nonumber \\ \\
k_{\rm act}^{(22)}(t_i,t_j) & = & L_{\rm c}^2 \, {\gamma ^{(G,G)}}({t_i},{t_j}), \\
k_{\rm act}^{(33)}(t_i,t_j) & = & B_{\rm c}^2 \, {\gamma ^{(G,G)}}({t_i},{t_j}) + B_{\rm
  r}^2 \, {\gamma ^{(dG,dG)}}({t_i},{t_j}) \nonumber \\ &&+ \, {B_{\rm{c}}}\,{B_{\rm{r}}}\left( {\,{\gamma ^{(G,dG)}}({t_i},{t_j})
     + {\gamma ^{(dG,G)}}({t_i},{t_j})} \right), \nonumber \\ \\
k_{\rm act}^{(12)}(t_i,t_j) & = & V_{\rm c} \, L_{\rm c} \, {\gamma ^{(G,G)}}({t_i},{t_j})
+ V_{\rm r} \, L_{\rm c} \, {\gamma ^{(G,dG)}}({t_i},{t_j}), \nonumber \\
 \\
k_{\rm act}^{(13)}(t_i,t_j) & = & V_{\rm c} \, B_{\rm c} \, {\gamma ^{(G,G)}}({t_i},{t_j})
+ V_{\rm r} \, B_{\rm r} \, {\gamma ^{(dG,dG)}}({t_i},{t_j}) \nonumber \\ &&+ V_{\rm
  c} \, B_{\rm r} \, {\gamma ^{(G,dG)}}({t_i},{t_j}) + V_{\rm r} \, B_{\rm c} \, {\gamma ^{(dG,G)}}({t_i},{t_j}), \nonumber \\
  \\ 
k_{\rm act}^{(23)}(t_i,t_j) & = & L_{\rm c} \, B_{\rm c} \, {\gamma ^{(G,G)}}({t_i},{t_j})
+ L_{\rm c} B_{\rm r} \, {\gamma ^{(G,dG)}}({t_i},{t_j}). \nonumber \\
\end{eqnarray}
}
The final covariance matrix over the new input space is formed as the following block matrix:
\begin{equation}
{{\mathbf{K}}_{{\text{act}}}} = \left( {\begin{array}{*{20}{c}}
  {{\mathbf{k}}_{{\text{act}}}^{(11)}}&{{\mathbf{k}}_{{\text{act}}}^{(12)}}&{{\mathbf{k}}_{{\text{act}}}^{(13)}} \\ 
  {{\mathbf{k}}_{{\text{act}}}^{(21)}}&{{\mathbf{k}}_{{\text{act}}}^{(22)}}&{{\mathbf{k}}_{{\text{act}}}^{(23)}} \\ 
  {{\mathbf{k}}_{{\text{act}}}^{(31)}}&{{\mathbf{k}}_{{\text{act}}}^{(32)}}&{{\mathbf{k}}_{{\text{act}}}^{(33)}} 
\end{array}} \right).
\end{equation}
Provided $\gamma$ is a valid covariance function, ${{\mathbf{K}}_{{\text{act}}}}$ is guaranteed to be a valid covariance matrix: in particular, it will be symmetric and positive semi-definite. In practice, then, it is not necessary to compute the blocks ${{\mathbf{k}}_{{\text{act}}}^{(21)}}$, ${{\mathbf{k}}_{{\text{act}}}^{(31)}}$, and ${{\mathbf{k}}_{{\text{act}}}^{(32)}}$ explicitly.

Thus, equipped with input vector $\mathbf{x}$, output vector $\mathbf{y}$, and covariance matrix ${\mathbf{K}_\text{act}}$, we are in a position to use the standard machinery of Gaussian process inference to model all three time series jointly as manifestations of a single underlying Gaussian process. However, it remains for us to choose the form of the covariance function, $\gamma$, defining this underlying stochastic process.
\subsection{Choice of latent covariance function}\label{sec:covFunc}
We now seek the simplest form for the covariance kernel function, {$\gamma$}, that satisfies our beliefs and ignorance about the activity process we are modelling.
\subsubsection{Squared-exponential covariance}
For many time series, we expect the informativeness of past observations in explaining current data to be a function of how long ago the past observations were made (for example, we might expect two observations made minutes apart to have similar values, but would not necessarily expect any connection between observations made months apart); if this is the case, we can restrict our consideration to \emph{stationary} covariance functions which are dependent solely on $\left| t - t' \right|$. Such covariance functions can be represented as the Fourier transform of a normalised probability density function \citep[via Bochner's Theorem][]{rasmussen2006}, which can be interpreted as the spectral density of the process.

The most widely-used covariance function of this class -- indeed, probably the most widely-used covariance function in GP inference -- is the squared-exponential or Gaussian kernel function, given by
\begin{equation}
\label{eq:SEcov_simple}
\gamma ^{(G,G)}(t,t') = {\beta ^2}\exp \left[ { - \frac{{{{(t - t')}^2}}}{{2{\lambda ^2}}}} \right].
\end{equation}
The hyper-parameter $\beta$ governs the output scale (gain/amplitude) of our function, and $\lambda$ controls the time scale. In our problem the amplitude parameter $\beta$ is not strictly necessary since $G(t)$ is never directly observed, and $\beta$ is absorbed by the model parameters $V_\textrm{c}$, $V_\textrm{r}$, $L_\textrm{c}$, $B_\textrm{c}$, and $B_\textrm{r}$, as in Equations \ref{eq:FFprime1}--\ref{eq:bis}.  

The popularity of this covariance function stems not only from its simplicity, but also the fact that it is infinitely differentiable, meaning that a GP with this covariance function will generate functions with no sharp discontinuities.

The squared-exponential covariance function is usually flexible enough to model a wide range of smoothly-varying stochastic processes. However, as discussed in Section~\ref{sec:intro}, we expect RV data and ancillary time series to show evidence for activity-induced variations on up to three distinct time-scales: years (magnetic activity cycle), weeks/months (rotation), and days (oscillation, granulation).  Thus it makes sense to generalise the covariance function in equation~\eqref{eq:SEcov_simple} to explicitly allow for multiple evolutionary time-scales:
\begin{equation}\label{eq:SEcov_gen}
\gamma _{se}^{(G,G)}(t,t') = \sum\limits_{i = 1}^N {{\beta _i^2}\exp \left[ { - \frac{{{{(t - t')}^2}}}{{2\lambda _i^2}}} \right]};
\end{equation}
here, the $\beta_i$'s control the relative amplitude of the $N\ge1$ components, each with evolutionary time-scale $\lambda_i$.

We initially experimented with $N=3$ separate additive terms, to describe possible variations on the short, medium, and long time-scales discussed in Section \ref{sec:nuisance}. \textcolor{black}{These tests were carried out using both synthetic data (see Section \ref{sec:tests-GP}) that by design incorporated variations on multiple different time-scales, as well as real datasets including the Gl~15~A and \aCB~datasets considered in Sections \ref{sec:tests-howard} and \ref{sec:tests-alphaCen}.} However, we rapidly found that the longer time-scale term was \textcolor{black}{invariably unnecessary}: any reasonable choice for the medium time-scale term had to be fairly flexible, since rotational signals generally evolve from year to year, and so the longer term variations can be captured by the medium time-scale term. On the other hand, we found a separate short-time scale term was generally required: without it, the residuals in all time series considered often exhibited obvious correlations.

Appendix \ref{sec:SEappendix} gives expressions for the covariance between observations of $G$ and its derivative $\dot{G}$, and between two observations of $\dot{G}$, for the case when $G$ is defined by a squared-exponential covariance function as in equation \eqref{eq:SEcov_gen}.
\subsubsection{Quasi-periodic covariance}
While the squared-exponential covariance function given in the previous section is both flexible and an obvious starting point when not given much information about the stochastic process we are modelling, we can do better. In particular, we have a physical reason to expect a degree of periodicity in the activity signals, since they are modulated by the periodic rotation of the star. We therefore consider the following quasi-periodic covariance function for the latent process $G(t)$ -- as previously considered by A12 to model {observed variations} in the Sun's total irradiance, and by \citet{haywood2014} to model correlated noise in RV residuals following an application of the $FF'$ method -- formed by multiplying\footnote{As described in \citet{rasmussen2006}, a valid covariance function under any arbitrary (smooth) map remains a valid covariance function. Valid covariance functions can also be constructed by adding or multiplying simpler covariance functions.} a stationary kernel with a periodic one:
\begin{equation}
\label{eq:cov_qp}
 {\gamma _{qp}^{(G,G)}(t,t')} = \exp \left\{ - \frac{\sin^2 \left[\pi (t-t')/P\right]}{2 \lambda_{\rm p}^2} 
 - \frac{(t-t')^2}{2\lambda_{\rm e}^2} \right\},
\end{equation}
where $P$ and $\lambda_{\rm p}$ correspond to the period and length scale of the periodic component of the variations, and $\lambda_{\rm e}$ is an evolutionary time-scale. While $\lambda_{\rm e}$ has units of time, $\lambda_{\rm p}$ is dimensionless, as it is relative to $P$. Once again there is no need to introduce an amplitude parameter, as this will be controlled for various time series by the model parameters $V_\textrm{c}$, $V_\textrm{r}$, $L_\textrm{c}$, $B_\textrm{c}$, and $B_\textrm{r}$, as in Equations \ref{eq:FFprime1}--\ref{eq:bis}.  

\begin{figure*}
\begin{center}
\includegraphics[width=\textwidth]{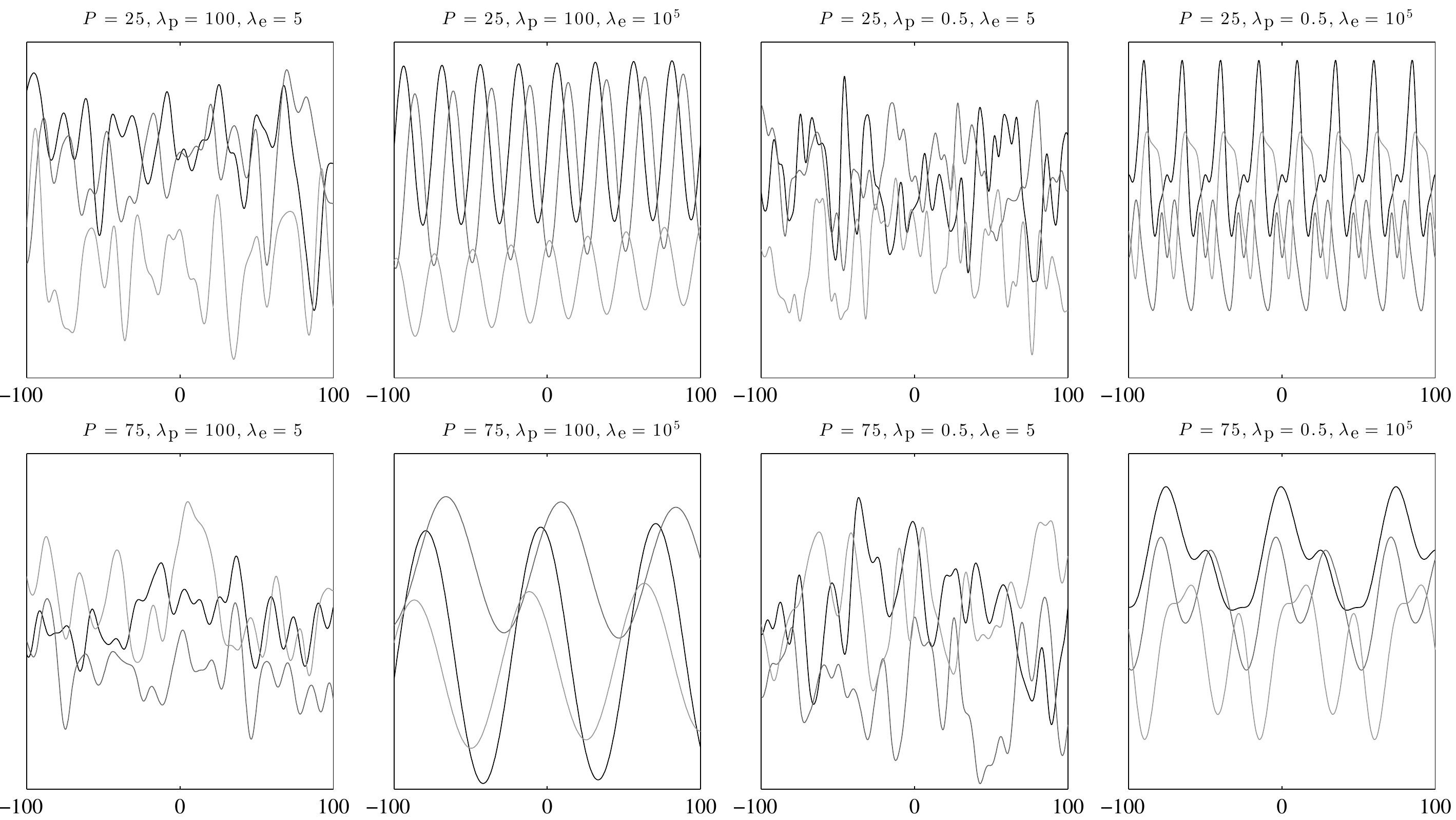}
\caption[]{Examples of functions drawn from a Gaussian process with a quasi-periodic covariance function, as defined in Equation \ref{eq:cov_qp}, in each panel with different hyper-parameter values (three function draws per panel). The hyper-parameter $P$ sets a characteristic period for the function's variations, $\lambda_P$ controls the extent to which the function varies within one period, and $\lambda_e$ controls the time-scale over which the (quasi-)periodic component evolves (variations become strictly periodic as $\lambda_e\to\infty$). In all cases, the function amplitudes are arbitrary, and a constant mean function (offset) is assumed.}
\label{fig:QP_example}
\end{center}
\end{figure*}

Examples of functions drawn from a Gaussian process with this quasi-periodic covariance function are given in Fig.\ \ref{fig:QP_example}.

\textcolor{black}{A word is in order about the interpretation of the hyper-parameters. While they could be related to spot persistence lifetimes, complexities of spot configurations, autocorrelation function decay time-scales, etc., such direct interpretations are rarely straightforward or well-grounded theoretically, especially because of degeneracies between the hyper-parameters. In the case of modelling differential rotation and/or spot evolution, for example, the hyper-parameter $P$ might become almost irrelevant, with a large value of $\lambda_P$ and a very small value of $\lambda_e$ allowing for complex, rapidly evolving waveforms without any obvious characteristic periodicity. In future, an in-depth study of how the values of these hyper-parameters translate into physically-interpretable properties, for example using the large sample of Kepler light curves considered by \cite{Aigrain2015} -- which in turn highlights the difficulty of distinguishing between differential rotation and spot evolution -- might be useful.}

In any event, the quasi-periodic covariance kernel function defined in equation \eqref{eq:cov_qp} is physically-motivated (\textcolor{black}{albeit not always easy to interpret straightforwardly}) and has only three hyper-parameters; moreover, in tests with both real and simulated data, we found it to produce as good or better results than when using a squared-exponential covariance function with a similar number of hyper-parameters. Therefore in all subsequent discussion the quasi-periodic covariance function will be assumed by default. (We do however discuss the squared-exponential covariance function because it is conceivable that there will be cases, though not considered here, where it might be a more appropriate choice than the quasi-periodic covariance; alternatively, it might be appropriate to generalise the function to include more than one quasi-periodic component.)

Appendix \ref{sec:QPappendix} gives expressions for the covariance between observations of $G$ and its derivative $\dot{G}$, and between two observations of $\dot{G}$, for the case when $G$ is defined by a squared-exponential covariance function as in equation \eqref{eq:cov_qp}.
\subsubsection{Other possible covariance functions}
There exists a wide variety of functions that can be used to specify the correlation between pairs of inputs, and thence to generate a covariance matrix over a set of observations and predicants. These functions can further be combined and modified in a multitude of ways, giving one great flexibility in how functions are modelled. In the preceding subsections, we considered two simple such functions that were deemed adequate for the cases considered in this paper; however, it would certainly be possible to replace these and use, within our same framework, other covariance functions \citep[e.g.][]{rasmussen2006}. 

For example, as an alternative to using a sum of squared-exponential kernels, one might use a rational-quadratic kernel, given by:
\begin{equation}
	\gamma _{rq}^{(G,G)}(t,t') = {\beta ^2}{\left( {1 + \frac{{{{(t - t')}^2}}}{{\alpha {\lambda ^2}}}} \right)^{ - \alpha }};
\end{equation}
it may be shown that this is equivalent to a scale mixture of squared-exponential kernels with different length scales, with the latter being distributed according to a Beta distribution with parameters $\alpha$ and ${\lambda ^{ - 2}}$.

On the other hand, if one has reason to expect that the latent process being modelled is not as smooth as functions drawn from a GP with squared-exponential kernel, one could consider the Mat\'ern class of covariance functions, defined by:
\begin{equation}
\gamma _M^{(G,G)}(t,t') = \frac{{{\beta ^2}}}{{\Gamma (\nu ){2^{\nu  - 1}}}}   {\left( {\sqrt{2\nu}   \frac{{\left| {t - t'} \right|}}{\lambda }} \right)}^{\nu}  {\mathbb{B}_\nu }\left( {\sqrt{2\nu}   \frac{{\left| {t - t'} \right|}}{\lambda }} \right),
\end{equation}
where $\beta$ is an output scale, $\lambda$ is the input scale, $\Gamma()$ is the standard Gamma function such that $\Gamma(\nu)=(\nu-1)!$, and $\mathbb{B}()$ is the modified Bessel function of second order. The hyperparameter $\nu$ controls the degree of differentiability of the resultant functions modelled by a GP with Mat\'ern covariance function, such that they are only $k$-times differentiable if $\nu>k$. For example, taking $\nu=\tfrac{5}{2}$ leads to twice-differentiable functions, and as $\nu\to\infty$, the Mat\'ern kernel becomes the squared-exponential one. Though a Mat\'ern covariance function with $\nu<\infty$ might be deemed more physically realistic than a squared-exponential one, its use does of course come with the penalty of some additional mathematical and computational complexity. For the purposes of the present work, then, we did not carry out any tests with these more sophisticated covariance kernels; their use might, however, feature in future work.\footnote{Initial tests with the rational-quadratic kernel function led to very similar results as when using the squared-exponential kernel. A rigorous approach to choosing between models with different kernel functions would require computation of Bayes factors for all models considered, an extension to this work we defer to future consideration.}
\subsection{Summary and appraisal of the framework}
\begin{figure}
\begin{center}
\includegraphics[width=\columnwidth]{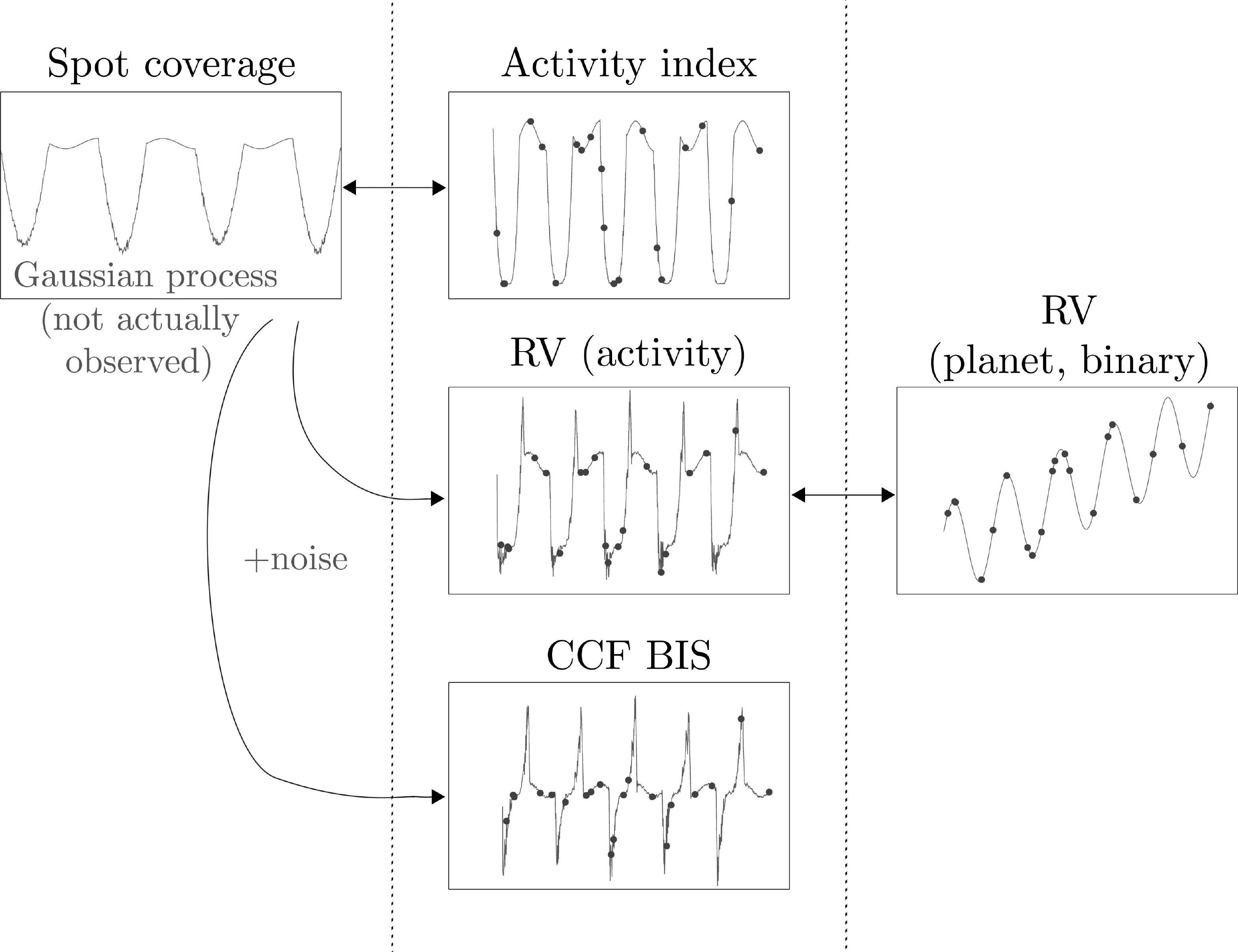}
\caption[]{Simplified, schematic sketch of the our GP-based scheme for the joint modelling of an RV time series with ancillary activity diagnostics. The unobserved, stochastic process giving rise to activity-related RV variations (left panel) is assumed to be describable by a GP. The way in which this propagates into observable time series (middle panels) is computed analytically, using a simple physical model, and so all observed time series can be modelled jointly. A noise model and deterministic physical component (right panel, e.g.\ Keplerian model to describe exoplanet-induced RV variations) completes the basic framework.}
\label{fig:GP_fitting_scheme}
\end{center}
\end{figure}

The key aspects of the above framework are sketched schematically in \mbox{Fig.\ \ref{fig:GP_fitting_scheme}}. Though it may appear quite mathematically complex, especially to those unfamiliar with GP regression, its underlying ideas are quite simple. To summarise:
\begin{enumerate}
\item we assume that the underlying stochastic process giving rise to activity signals in observables (RVs and ancillary time series) can be described by a GP, with suitably-chosen covariance function;
\item we can then use physically-motivated or empirical models to provide analytical links between this GP and available observables;
\item finally, with the addition of noise and deterministic components (e.g.\ dynamical effects for the RVs), all observables can be modelled jointly as GPs, with the ancillary time series serving to constrain any activity component of the RVs.
\end{enumerate}

Our framework offers a number of advantages over existing approaches. Firstly, modelling all available activity-sensitive time series (line widths, asymmetries, chromospheric activity indices etc.) jointly with RVs should allow tighter constraints to be placed on activity signals in RVs, compared to exploiting only simple correlations between RVs and (typically) one of these other time series. Secondly, our general framework is flexible in that it uses Gaussian process draws (and derivatives thereof) as basis functions to model available time series, rather than e.g.\ sinusoids or other simple parametric models, the inappropriate use of which could lead to the inadvertent introduction of correlated signals into model residuals; in the case our GP framework, the properties of basis functions are informed in a more principled way by the observables. Thirdly, our framework facilitates smooth interpolation between observables, as well as extrapolation to future times (prediction). Lastly, being based on GPs, the entire framework can be accommodated very naturally within the broader framework of Bayesian inference, allowing uncertainties to be handled in a principled way: overly-complex models are automatically penalised, nuisance parameters can be marginalised over, and rigorous model comparisons can be performed\ \citep{rasmussen2006}.

A possible disadvantage of our framework is that, in its current form, it can only use the latent `activity' GP and linear transformations thereof (including derivatives) to generate the activity signals in observables; this is sufficient for linking e.g.\ RVs and photometry via the  $FF'$ method, but the latter is only a first-order model. Allowing for more general relationships between the latent Gaussian process and the observables might be possible, although it is not immediately clear that doing so would result in a computationally- or mathematically-tractable framework. Another potential disadvantage is that `vanilla' Gaussian-process regression is not well-suited to modelling large datasets ($N\gg1000$ observations): GP regression hinges on matrix inversion, with na\"ive inversion algorithms scaling as $N^3$. Fortunately there do exist a number of approximation techniques which hold great promise for allowing efficient GP regression on larger datasets \citep{lazaro2010,chalupka2012,ambi2014}. In the case of time-series analysis, there exist very efficient $\mathcal{O}(N)$ approaches for Gaussian process regression based on approximate re-interpretation as filters \citep{reece2010e,sarkka2013,reece2014}.

It should be noted that this is not the first instance of GPs being used to model temporally-correlated stellar nuisance signals: \citet{hou2014thesis} presented a framework for using GPs to model stochastic stellar oscillations in RV data, while \citet{barclay2014arxiv} used GPs to model photometric granulation signals in a Kepler-91 dataset. The only published instances (to our knowledge) of GPs being used specifically to model stellar \emph{activity} signals are the work of \citet{haywood2014}, who used a GP to mitigate the effects of \textcolor{black}{active regions without photometric counterparts} in the CoRoT-7 system, and the work of \citet{grunblatt2015}, who used a GP to model activity in Kepler-78 datasets. In both cases, quasi-periodic GPs were trained on photometric data, and a GP model with the same covariance properties then used to model available RV data. The framework we have presented may be seen as a more general and flexible version of the aforesaid approach, in the sense that our framework allows \emph{all} available activity-sensitive time series to be modelled jointly, and does not necessarily require simultaneous RV and photometry data. \textcolor{black}{When both RV and photometric data are available, the $FF'$ formalism is built directly into the modelling; unlike the $FF'$ method, however, our framework would not require all active regions to have photometric signatures in order to constrain their RV contributions, provided at least one other activity-sensitive time series (e.g.\ \bis) were available.}

\section{Tests and applications}\label{sec:tests}
\subsection{Simulated data using na\"ive physical model}\label{sec:tests-GP}
As a first test of our GP framework, we used a number of different functions (e.g.\ polynomials; sinusoids; wavelets; function draws from GPs; etc.) to generate a range of synthetic time series to represent possible latent, activity-driving processes. \textcolor{black}{We included periodicities and correlations on many different time-scales, to test our models' ability to deal simultaneously with the three different time-scales for variations discussed in Section \ref{sec:nuisance}} (\textcolor{black}{as a simple example, a sine wave with evolving amplitude, added to a long-term parabolic trend)}. These time series were densely-sampled and noise-free. We then used the relations in equations \eqref{eq:phys-model-1}, \eqref{eq:phys-model-2}, and \eqref{eq:phys-model-3} to simulate \drv, \lrhk\ and \bis\ ``observables'', computing derivatives numerically using finite-difference approximation where applicable. In other words, we generated the observables we would obtain for arbitrary activity processes, assuming the physical model we used to link the activity process to observables corresponded exactly to reality.

Using a number of different covariance functions (including those discussed in Section \ref{sec:covFunc}), our general findings were the following.
\begin{enumerate}
\item By performing unconstrained optimisation of the free parameters in model likelihood functions $\mathcal{L}_{m,k}(\bmg{\theta},\bmg{\phi})$, all observed time series could be modelled very accurately, with fits that appeared ``exact'' upon visual inspection (though in fact with small residuals consistent with numerical error due to finite precision arithmetic, matrix inversion, derivative approximations, etc.).\footnote{Multi-modality did not seem to pose a significant problem with the likelihood functions in question: a single run of a simple Nelder-Mead (``downhill simplex'') algorithm often sufficed to locate globally-optimal solutions, though in a few cases the algorithm did benefit from multiple starting locations.}
\item In all such cases, the fitted model's latent process (GP) was also correspondingly close to the latent process used to generate the observables. In cases where the latent process used to generate the observables was itself a GP, the hyper-parameters of this latent process could be recovered accurately.
\item Imposing constraints on the hyper-parameters of chosen mean and covariance functions led to the expected changes in fits. For example, constraining the evolutionary time-scale $\lambda_e$ (for the quasi-periodic covariance kernel) to be longer than a certain time prevented physically unrealistic, short-time scale ``contortions'' in the fits. Similarly, forcing a finite period parameter $P$ (again in the quasi-periodic covariance kernel), with long evolutionary time-scale $\lambda_\textrm{e}$, led to poor fits when the simulated observables contained no actual periodicities.
\item When adding \textcolor{black}{white} Gaussian noise to the simulated observables, the quality of fits remained good -- as measured by a reduced chi-square statistic -- provided that the estimated amplitudes of the added noise were reasonable.
\end{enumerate}
In sum, all aspects of the modelling worked as expected; we established that the analytical aspects of (if not the physical assumptions underlying) our framework were watertight, and that when this framework was implemented numerically, we could indeed jointly model multiple time series generated by an unobserved stochastic process and its derivative.

\subsection{Simulated data from SOAP 2.0}\label{sec:tests-SOAP}
\begin{figure*}
\begin{center}
\includegraphics[width=\textwidth]{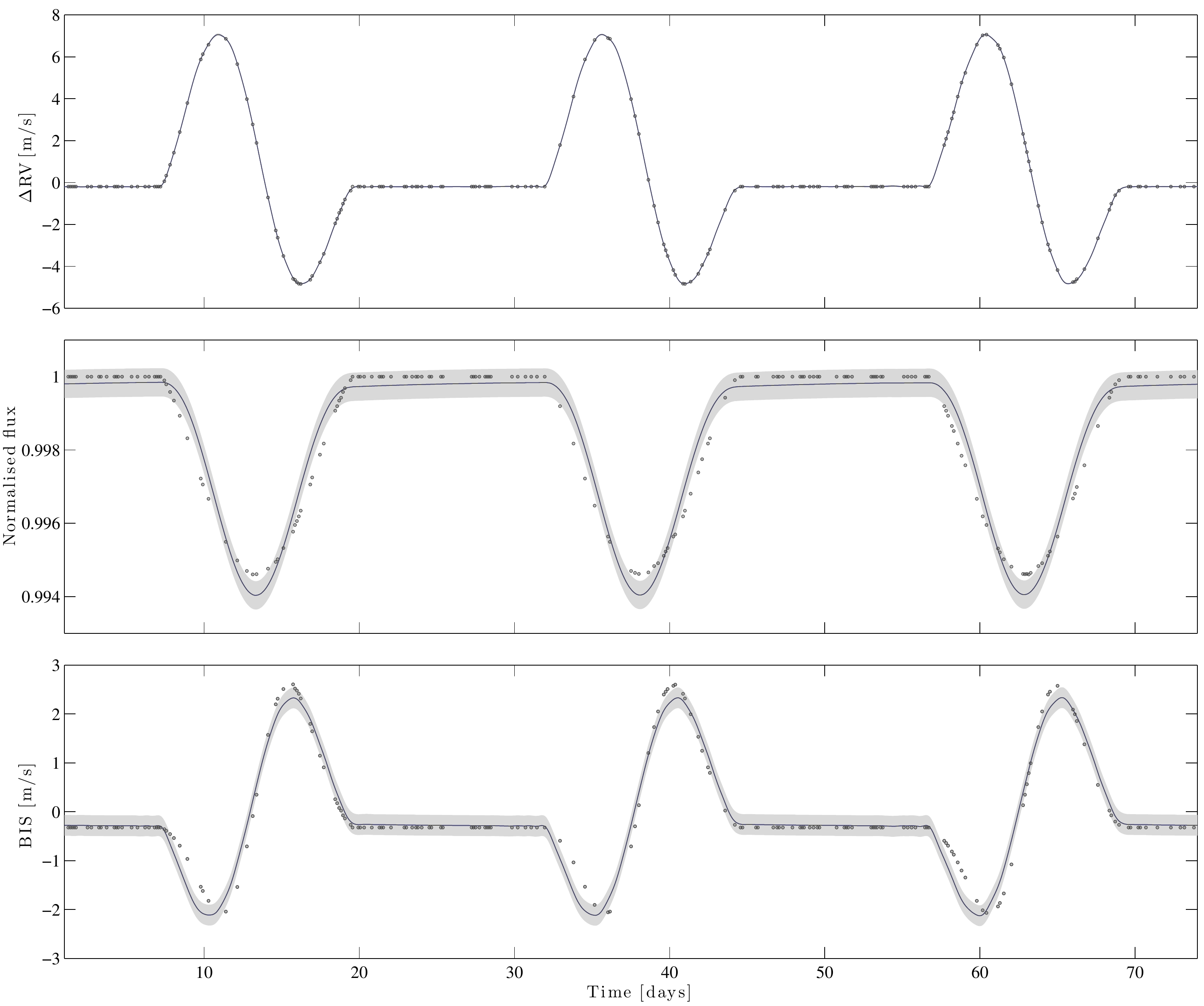}
\caption[]{GP model MAP fit to noise-free \mbox{SOAP 2.0} data, based on a simulation of a single rotating spot at latitude $\phi=45^\circ$, and radius $10\%$ of the stellar radius. The dots indicate the simulated observations; the solid lines are model posterior means, and the shaded regions denote $\pm\sigma$ posterior uncertainty. The posterior uncertainty in each time series is directly related to the amount of additive noise ($\sigma_{l_i}$ terms) favoured in the MAP model.}
\label{fig:SOAP_001a}
\end{center}
\end{figure*}
\begin{figure*}
\begin{center}
\includegraphics[width=\textwidth]{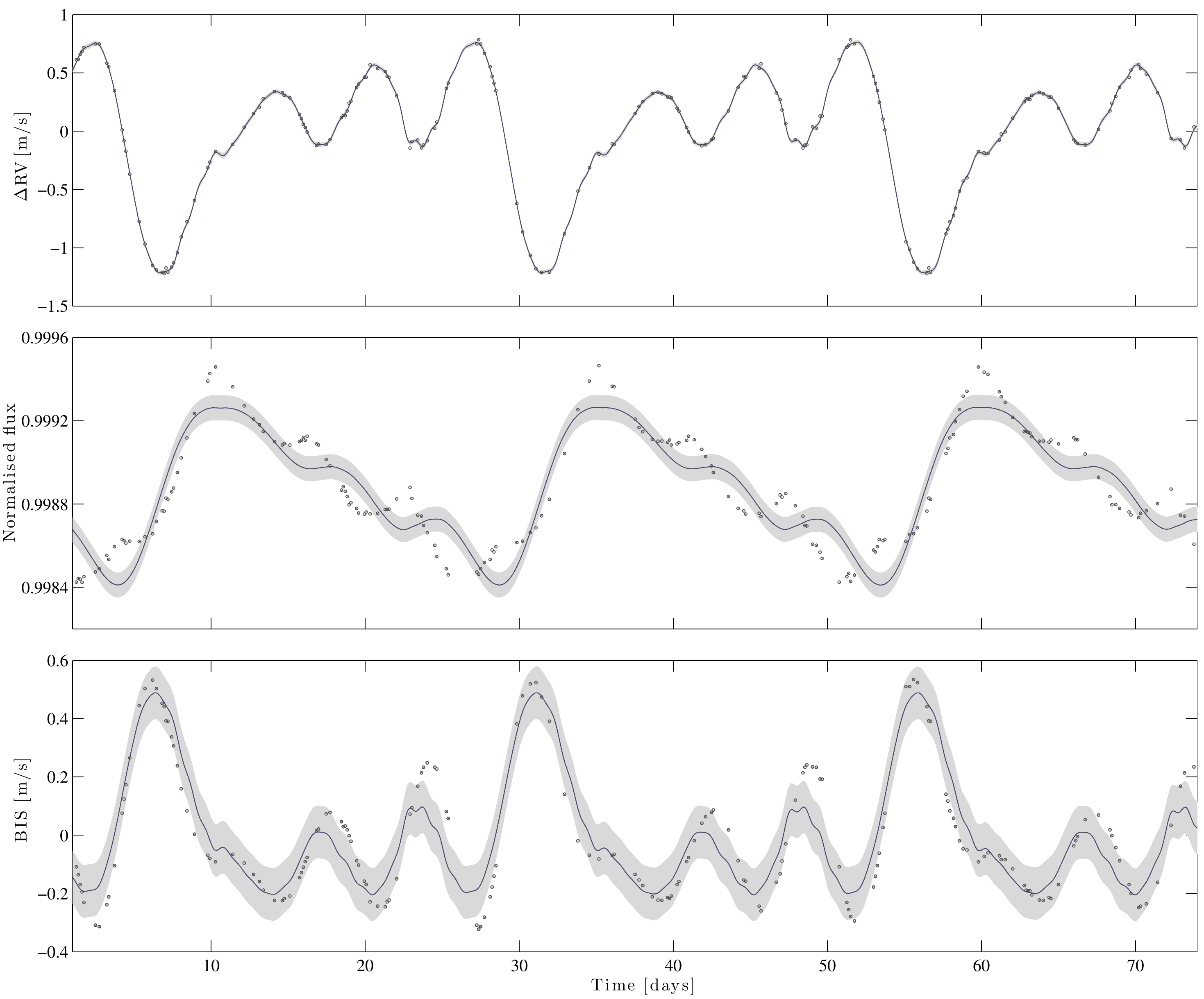}
\caption[]{GP model MAP fit to noise-free \mbox{SOAP 2.0} data, based on a simulation of four rotating spots (each with radius $5\%$ of the stellar radius) at latitudes $\phi=0^\circ,30^\circ,45^\circ,60^\circ$, equally spaced in longitude. The dots indicate the simulated observations; the solid lines are model posterior means, and the shaded regions denote $\pm\sigma$ posterior uncertainty. The posterior uncertainty in each time series is directly related to the amount of additive noise ($\sigma_{l_i}$ terms) favoured in the MAP model.}
\label{fig:SOAP_001b}
\end{center}
\end{figure*}

SOAP (Spot Oscillation and Planet) 2.0 is a software package developed by \citet{dumusque2014soap2} that simulates the effect of stellar spots and plages on radial velocity and photometry; it extends an older code, SOAP, developed by \citet{boisse2012}. The tool is available online at \url{http://www.astro.up.pt/soap2}.

\begin{figure*}
\begin{center}
\includegraphics[width=\textwidth]{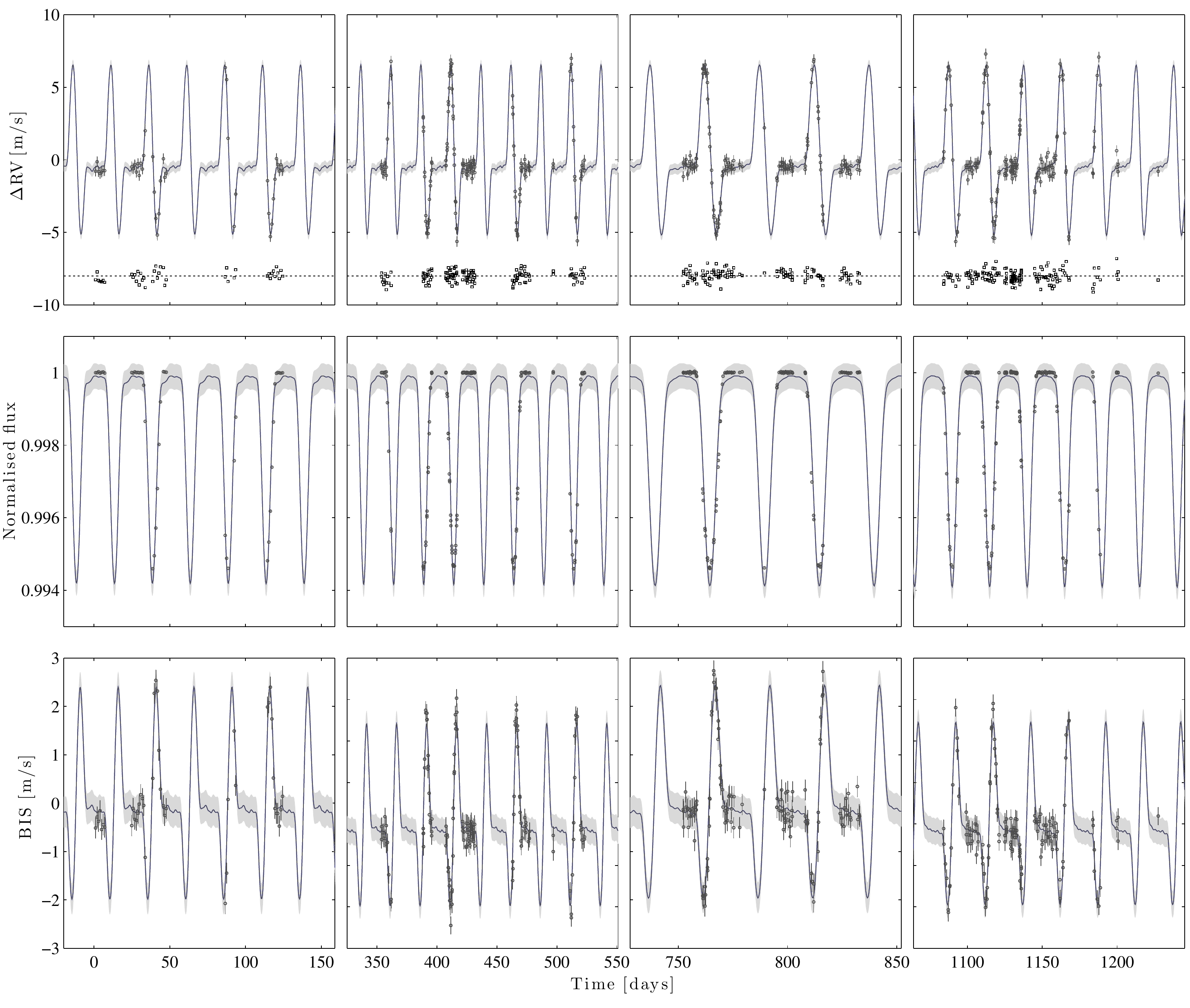}
\caption[]{GP model MAP fit to \mbox{SOAP 2.0} data, based on the same one-spot configuration as in Fig.\ \ref{fig:SOAP_001a}, but this time with time sampling and noise levels taken from a real HARPS dataset. Each time series is split into four panels, to avoid plotting large gaps where no data are present. Residuals in the \drv~time series are plotted below the simulated data and fitted model, but for the sake of clarity, with an arbitrary vertical offset from the main time series.}
\label{fig:SOAP_002a}
\end{center}
\end{figure*}
\begin{figure*}
\begin{center}
\includegraphics[width=\textwidth]{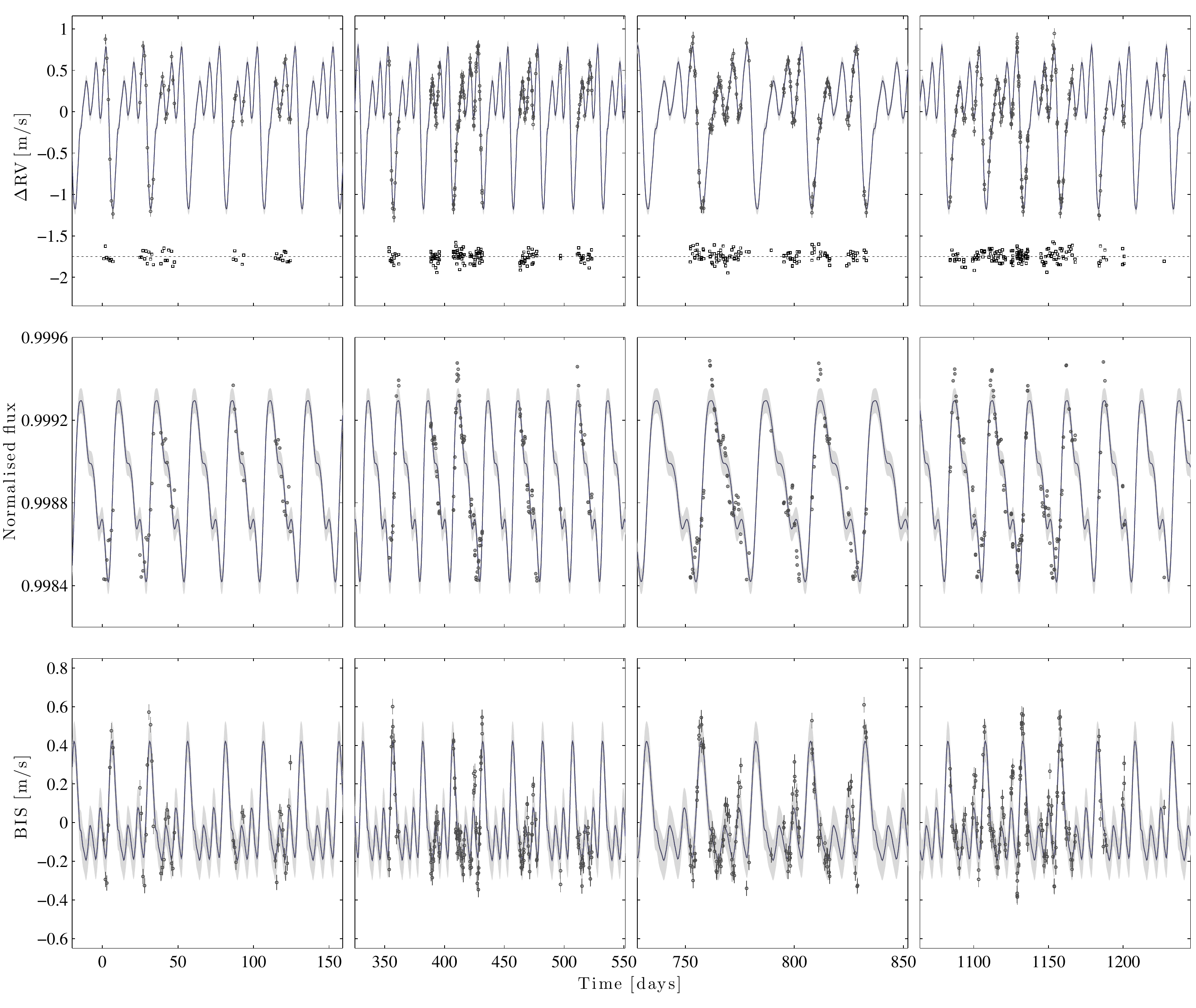}
\caption[]{GP model MAP fit to \mbox{SOAP 2.0} data, based on the same four-spot configuration as in Fig.\ \ref{fig:SOAP_001b}, but this time with time sampling and noise levels taken from a real HARPS dataset. Each time series is split into four panels, to avoid plotting large gaps where no data are present. Residuals in the \drv~time series are plotted below the simulated data and fitted model, but for the sake of clarity, with an arbitrary vertical offset from the main time series.}
\label{fig:SOAP_002b}
\end{center}
\end{figure*}

In brief, the original SOAP code works by dividing up a simulated stellar disk into tens of thousands of resolution elements, with a CCF (cross-correlation function, representing the typical spectral line) for each element then being modelled by a Gaussian. Each CCF is Doppler-shifted according to the projected rotational velocity, and weighted by a linear limb-darkening law. In this way, the non-spotted emission of the visible disk may be computed in photometry and in RV. SOAP then calculates the positions of rotating surface inhomogeneities, which are defined by their latitudes, longitudes, and sizes; the weighted CCF and flux for cells inside the inhomogeneities are added to (in the case of bright spots or plages) or removed from (in the case of dark spots) those of the non-spotted star. Finally, SOAP delivers the integrated spectral line, the flux, the RV, and the BIS as a function of stellar rotational phase.

SOAP 2.0 extends SOAP by considering, additionally, the inhibition of convective blueshift inside active regions, the limb brightening effect of plages, a quadratic limb darkening law, a realistic active region contrast, and the resolution of the spectrograph used for the (simulated) observations.

A number of stellar, spot, and spectral parameters can be provided as inputs to the SOAP 2.0 code. In particular, the code allows up to four active regions (each specified by a size, brightness, latitude and longitude) to be included in the simulation.

The SOAP 2.0 code does incorporate simplifications in its modelling: for example, the observed spectrum of a plage is considered to be the same as that of a spot, and some of the stellar physics included in the modelling is informed only by Solar photosphere and spot spectra, which are assumed to be representative of all quiet photosphere regions and active regions. \textcolor{black}{It also does not include long-term magnetic activity cycles in its simulations.} Its output nevertheless provides a far more realistic set of test cases than those considered in Section \ref{sec:tests-GP} -- especially because it does not impose any \emph{ad hoc} constraints on the functional relationship between photometric, RV and BIS time series. Moreover, \citet{dumusque2014soap2} showed that \mbox{SOAP 2.0} manages to reproduce the activity-induced variation on the early-K dwarfs \mbox{HD 189733} and \aCB; modelling the activity-induced variation on the latter star with our new GP framework is considered in Section \ref{sec:tests-alphaCen}.

Given the wide range of possible outputs we could generate from \mbox{SOAP 2.0} (and then modify in various ways), we adopted the following scheme in order to make a systematic study of how our GP modelling framework could be applied to realistic data.
\begin{enumerate}
\item Generate noise-free, densely-sampled outputs for a number of different spot configurations, and perform fits to these outputs (modelling photometric flux in place of \lrhk, i.e.\ using the $FF'$~model; see Section \ref{sec:physical-model}). Based on these results, select a few qualitatively different configurations -- including the configuration for which the poorest fits are obtained -- for further study.
\item Study the residuals from the fits in the selected noise-free cases, and examine how much additive white noise -- i.e.\ the $\sigma_{l_i}$ terms in Equation \eqref{eq:whitenoise} -- should be allowed in the various time series to ensure that activity is accurately filtered from RVs.
\item Repeat the fitting as above (now allowing for additive white noise in the GP model), but this time modifying the SOAP code's output to include observational noise, realistic time sampling, etc. 
\item Repeat the fitting as above, but with the injection of a Keplerian signal into the RVs, and study whether the Keplerian signal can be disentangled from the activity signal.
\end{enumerate}

It should be emphasised that although this resembles a principled approach to studying our modelling framework using the \mbox{SOAP 2.0} tool, we did not attempt to make a comprehensive study of (amongst other things) the inter-related effects of different levels of observational noise, different time samplings, different Keplerian amplitudes, periods or phases, etc. Apart from being beyond the scope of this work, such an all-encompassing study would likely have been misguided, given that both our model and the \mbox{SOAP 2.0} tool incorporates simplifying assumptions which are not easily quantified. As such, discrepancies between SOAP output and our GP models would not necessarily represent shortcomings of our model; conversely, good agreement between SOAP output and our model would not necessarily imply equally good agreement could be expected with real observational data. Rather, the purpose of these tests was to ascertain whether our framework would work at all with fairly realistic simulated data, to understand how the $\sigma_{l_i}$ terms in our model should be constrained, and to get a conservative though quantifiable estimate of the possible shortcomings of our model.

We found that by allowing proportionately more additive noise (the $\sigma_{l_i}$ terms in Equation \ref{eq:whitenoise}) in the ancillary time series than the RV time series -- in effect, attaching less weight in the fitting to these time series, though still using them to constrain the latent process' hyper-parameters -- we were generally able to obtain RV residuals that resembled white Gaussian noise. In other words, we were only interested in modelling activity-induced variations in \drv~accurately, and we could achieve this goal provided we ``loosened'' the assumed relationships between \drv~and ancillary, activity-sensitive time series (see Section \ref{sec:GP-model}). When we did not allow for any white-noise in any of the time series, or set equal upper bounds on the white noise in all time series (effectively giving equal weight to all time series), we generally ended up with better fits to the ancillary time series, poorer fits for \drv, and correlated residuals in all time series. 

In particular, we deduced the following approximate upper bounds to allow on the $\sigma_{l_i}$ terms, if we wanted to ensure white Gaussian RV residuals:
\begin{itemize}
\item $\sigma_{\Delta\textrm{RV}}$: up to $5\%$ of $\textrm{rms}(\Delta\textrm{RV})$;
\item $\sigma_\textrm{flux}$: up to $10\%$ of $\textrm{rms}(\textrm{flux})$;
\item $\sigma_\textrm{BIS}$: up to $20\%$ of $\textrm{rms}(\textrm{BIS})$.
\end{itemize}
Constraining the additive white noise in terms of the root mean square (rms) variability of a given time series reflects the fact that there are certain features in these time series that our model cannot capture accurately (e.g.\ an inflection point in the flux time series, when none exists in \drv), regardless of the details of the scaling, offsets, etc.\ of the time series in question. These constraints are simply upper bounds on $\sigma_{l_i}$, and serve to quantify the extent to which our framework fails to model all three time series simultaneously; in practice, the values of $\sigma_{l_i}$ will usually be smaller than these upper bounds, and will be determined by optimisation or marginalisation, as for any of the other free parameters in our framework.

\begin{figure*}
\begin{center}
\includegraphics[width=\textwidth]{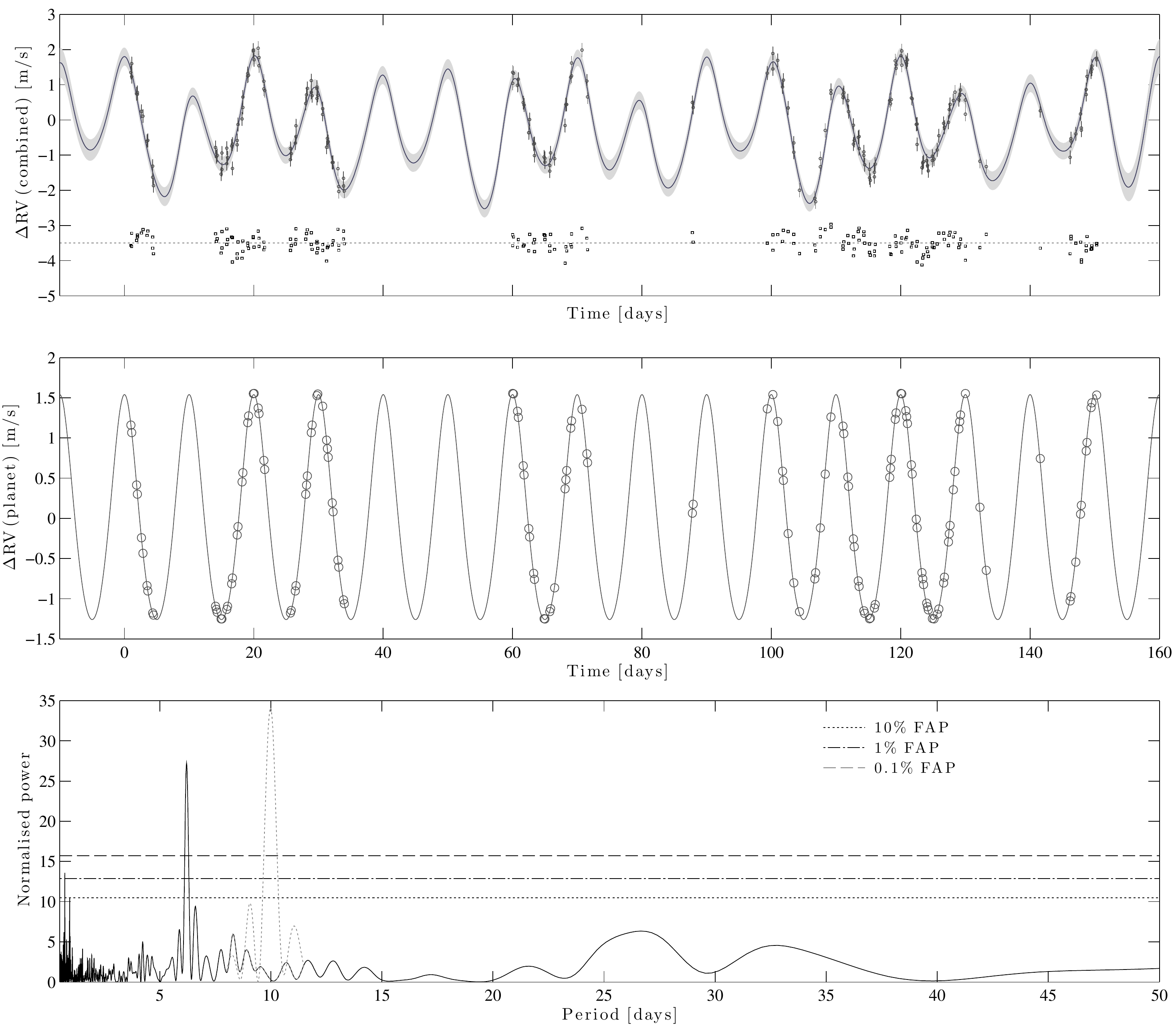}
\caption[]{GP model MAP fit to \mbox{SOAP 2.0} data, based on the same four-spot configuration as in Fig.\ \ref{fig:SOAP_001b}, with time sampling and noise levels taken from one season of a real HARPS dataset, plus an injected Keplerian signal. Here the injected signal has an amplitude comparable to, but period different from, the rotationally-modulated activity signals. The top panel shows the model fit to the \drv~time series, including residuals (fits to ancillary time series not shown here); the middle panel shows the Keplerian component of the fit (open circles represent injected signal \emph{before} noise was added); and the bottom panel shows the normalised Lomb-Scargle periodogram of the \drv~residuals, along with false alarm probability thresholds. \textcolor{black}{The dotted grey peak in the periodogram around $10.0$~d indicates the significant excess power that remains in the \drv~residuals when performing a fit \emph{without} a planet (the rest of the power spectrum remains qualitatively similar, and for clarity is not included in full)}.}
\label{fig:SOAP_planet_A}
\end{center}
\end{figure*}
\begin{figure*}
\begin{center}
\includegraphics[width=\textwidth]{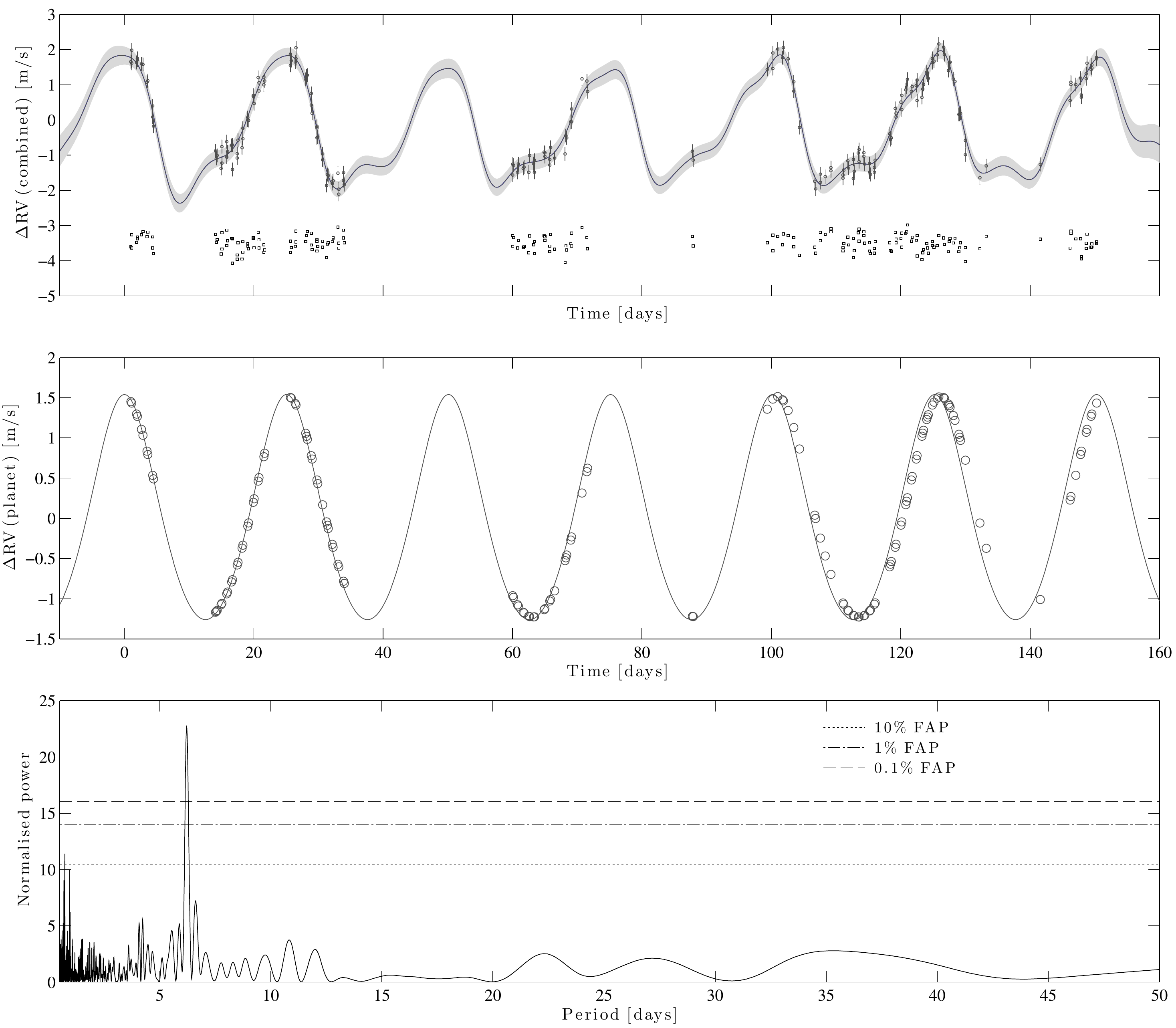}
\caption[]{GP model MAP fit to \mbox{SOAP 2.0} data, based on the same four-spot configuration as in Fig.\ \ref{fig:SOAP_001b}, with time sampling and noise levels taken from one season of a real HARPS dataset, plus an injected Keplerian signal. Here the injected signal has an amplitude comparable to, and period identical to, the rotationally-modulated activity signals. The top panel shows the model fit to the \drv~time series, including residuals (fits to ancillary time series not shown here); the middle panel shows the Keplerian component of the fit (open circles represent injected signal \emph{before} noise was added); and the bottom panel shows the normalised Lomb-Scargle periodogram of the \drv~residuals, along with false alarm probability thresholds.}
\label{fig:SOAP_planet_B}
\end{center}
\end{figure*}
\begin{figure*}
\begin{center}
\includegraphics[width=\textwidth]{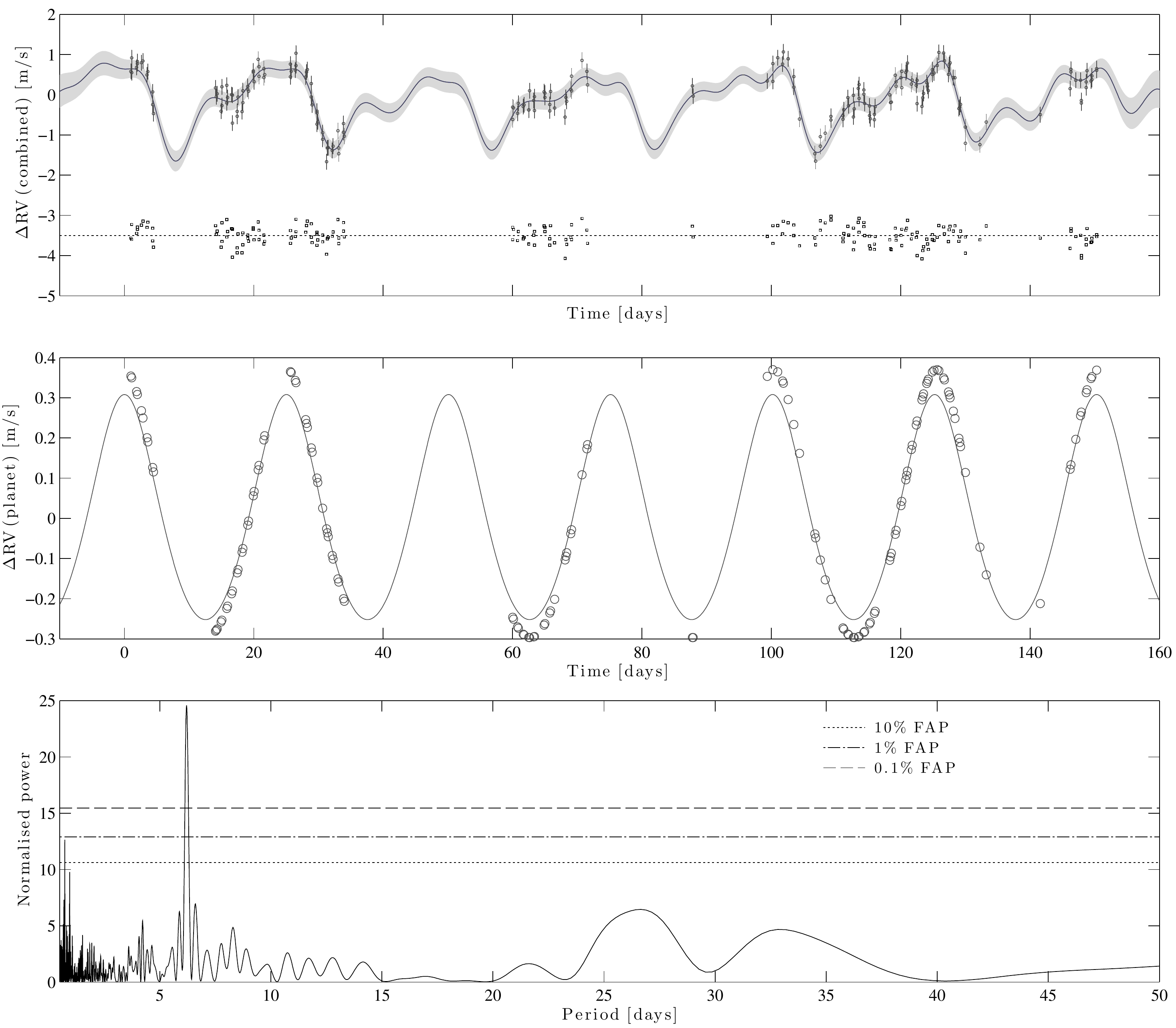}
\caption[]{GP model MAP fit to \mbox{SOAP 2.0} data, based on the same four-spot configuration as in Fig.\ \ref{fig:SOAP_001b}, with time sampling and noise levels taken from one season of a real HARPS dataset, plus an injected Keplerian signal. Here the injected signal has an amplitude five times smaller than, and period identical to, the rotationally-modulated activity signals. The top panel shows the model fit to the \drv~time series, including residuals (fits to ancillary time series not shown here); the middle panel shows the Keplerian component of the fit (open circles represent injected signal \emph{before} noise was added); and the bottom panel shows the normalised Lomb-Scargle periodogram of the \drv~residuals, along with false alarm probability thresholds.}
\label{fig:SOAP_planet_C}
\end{center}
\end{figure*}

To illustrate our findings, we consider the two representative configurations of a single spot (spot radius $10\%$ of the stellar radius) at latitude $\phi=45^\circ$, and of four equally-sized spots (each with radius $5\%$ of the stellar radius) at $\phi=0^\circ,30^\circ,45^\circ,60^\circ$, equally spaced in longitude. These configurations are representative of cases that were particularly easy and difficult, respectively, to model with our GP framework. In both cases, the observed star in the simulation had Solar parameters \citep[$5778$~K effective temperature, $5115$~K spot temperature, $25.05$~d rotation period, Solar limb darkening coefficients, etc., as per][]{dumusque2014soap2}, and the simulated instrument resolution (\num[group-separator={,}]{115000}) corresponded to a HARPS-like instrument. The observations we simulated were noise-free, and spanned three rotation periods (approximately $75$~days), with uniformly randomised temporal sampling. 

We \textcolor{black}{obtained MAP fits} of our GP model to the simulated data; we used a quasi-periodic covariance kernel, and placed non-informative priors on all free model parameters. Uniform priors were used for parameters with known scales (e.g.\ $P$, expected to correspond to a stellar rotation period; or $V_\textrm{r}$ and $V_\textrm{c}$, which set the amplitude for RV variations), while scale-invariant Jeffreys priors were used for all other parameters (e.g.\ $\lambda_\textrm{P}$, $\lambda_\textrm{e}$). \textcolor{black}{Typical computation time (on a modern laptop) was on the order of a few minutes to obtain a MAP solution, using a single run of a Nelder-Mead (downhill-simplex) algorithm. We established (probable) global optimality by making sure downhill-simplex runs from different starting points converged to the same optimum.}

Fig.\ \ref{fig:SOAP_001a} shows results for the `easy' (one-spot) configuration, while Fig.\ \ref{fig:SOAP_001b} shows results for the `difficult' (four-spot) configuration. In both cases, the \drv~time series could be modelled extremely accurately (rms of residuals on the order of $0.01$~\mps~in both cases). The ancillary time series were, necessarily, modelled less accurately -- especially in the case of the four-spot configuration, where the equatorial spot contributed the most problematic features to the ancillary time series. The magnitude of the $\sigma_{l_i}$ terms (additive white noise) used in the modelling is indicated by the size of the $\pm\sigma$ posterior uncertainties plotted in these figures.

We next considered using more realistic time sampling and adding noise to the simulated observations. To do so we used temporal sampling taken directly from the publicly-available \aCB~dataset from \citet{dumusque2012nature}, which featured nearly four years' worth of data, with gaps of several months between observing seasons. We also added \textcolor{black}{white} Gaussian noise to the SOAP~2.0 output: for each time series, the noise was scaled so to be the same fraction of the rms variation of the time series itself as was the corresponding \aCB~noise estimate.

Again we found that we could model \drv~very accurately: for both spot configurations, the rms of \drv~residuals was marginally smaller than the rms of the observational noise, indicating that our fits were limited only by our (simulated) noise floor. Figs.\ \ref{fig:SOAP_002a} and \ref{fig:SOAP_002b} show results for the `easy' and `difficult' spot configurations, as before. Despite the gaps in the observational coverage, the fitted models accurately recovered the structure of the periodic variations in all time series (\textcolor{black}{with the exception of a few outlier points, reflecting our model's inability to reproduce very sharp changes in one time series when another changes less sharply, since all time series are modelled with combinations of a single function and its first derivative}).  It is worth emphasising that when the data to be modelled include observational noise, the $\sigma_{l_i}$ terms in our model serve to account for both this noise \emph{and} any features in the ancillary time series that our model cannot accurately reproduce (we could artificially split the white-noise terms in the covariance matrix into `observational' and `model imperfection' components, although this would not affect in any way the practical implementation of our model).

Finally, having ascertained that our framework allowed activity-induced RV variations to be modelled accurately in isolation, we turned our attention to planet injection tests. We wanted to answer the question: \emph{Could we use our framework accurately to \emph{disentangle} activity-induced RV variations from other injected (non-activity) signals, e.g.\ planetary signals?} Our general finding was that, as expected, modelling the ancillary, activity-sensitive time series in conjunction with \drv~did indeed allow the activity component of \drv~to be very well constrained, but -- crucially -- without the non-activity RV signals being subsumed by the same GP model. Broadly speaking, we found that it became more and more difficult to disentangle injected Keplerian signals from activity signals as the amplitude of the injected signals grew smaller relative to the activity signals (and, of course, relative to observational noise). Having a Keplerian signal with a period very similar to that of the activity signal generally did not make disentangling activity and injected signal any more difficult, provided the amplitude of the Keplerian signal was not also much smaller than the activity signal.

To illustrate these findings, we present as examples three qualitatively-different Keplerian signals injected into the \drv~time series arising from the `difficult' four-spot configuration considered earlier. The activity-induced \drv~variations arising from this configuration had a semi-amplitude of $1.65$~\mps, and a period of $25.05$~days. We simulated observational noise using the \aCB~template as before, and we used time sampling corresponding to only one season of the \aCB~data, spanning approximately half a year (such that the simulated data sampled about 7 stellar rotations). The injected Keplerian signals we consider have the following properties (where $K$ is the RV semi-amplitude, and $P_K$ is the period of the variations; we use $P_K$ to distinguish this period from the period hyper-parameter $P$ in our quasi-periodic covariance kernel):
\begin{enumerate}
\item $K=1.4$~\mps, $P_K=10.0$~d;
\item $K=1.4$~\mps, $P_K=25.05$~d;
\item $K=0.28$~\mps, $P_K=25.05$~d.
\end{enumerate}
Other orbital parameters were identical in all three cases; in particular, eccentricity was fixed at $e=0.1$. Case (i) corresponds to an orbiting planet which induces an RV signal with amplitude comparable to the activity signal, and with a period that is not close to the stellar rotation period or any of its harmonics. Case (ii) adds the complication of the planet having a period very similar -- identical, in fact -- to the period of the rotational activity signal, while case (iii) goes one step further by representing a planetary signal with periodic identical to, but amplitude fives times smaller than, the activity signal.

Fits to data including these injected signals are presented in Figs.\ \ref{fig:SOAP_planet_A}, \ref{fig:SOAP_planet_B}, \ref{fig:SOAP_planet_C} (the Keplerian component was included in our GP model through its mean function; non-informative priors were placed on all Keplerian orbital parameters). MAP parameter estimates for selected Keplerian orbital elements in the planet injection tests, along with $\pm\sigma$ posterior uncertainties, are presented in Table \ref{tab:planet_fit}. All parameter inference was performed using the MultiNest nested-sampling algorithm \citep{multinest2008,multinest2009,multinest2013}, \textcolor{black}{with the GP hyper-parameters first fixed at their MAP values, as per the computational approximation motivated in \citet{gibson2012}. Typical computation time to obtain posterior distributions for physical parameters, using MultiNest on a modern laptop, and using MultiNest's default convergence criteria, was on the order of tens of minutes. Posteriors were generally unimodal (although in general, of course, this would depend on the model in question, the data sampling, etc.)}

In cases (i) and (ii), the fitted Keplerian amplitudes and periods agree with the true values to within a small fraction of a percent. In case (iii), the absolute errors in the fitted values are larger (the recovered amplitude, for example, is clearly a little smaller than the true amplitude), but \textcolor{black}{of course one could never expect perfect fits given an imperfect model, noise, and discrete time sampling. Nevertheless, it is reassuring that the true values do still lie within the $\pm2\sigma$ credible intervals around the MAP values}. 

In all cases, the rms of the \drv~residuals was on the order of $0.2$~\mps. The Lomb-Scargle periodogram\footnote{We computed normalised Lomb-Scargle periodograms based on the implementation described by \citet{Press2007}; false alarm probabilities were estimated by randomly permuting the original data, keeping observation times fixed.} of the \drv~residuals Figs.\ \ref{fig:SOAP_planet_A}, \ref{fig:SOAP_planet_B}, \ref{fig:SOAP_planet_C} all show significant peaks at $6.5$~d: we interpret this as the fourth harmonic of the activity signal, as this feature remained present in the \drv~residuals even when the injected Keplerian signal had a period that is not similar to the stellar rotation or any of its harmonics, e.g.\ as in case (i). The only other significant peaks in the periodogram of the residuals, \textcolor{black}{clustered around $1$~d, arise from the window function (time sampling) in the \aCB~dataset.}

\textcolor{black}{In cases (i), when we did not include a Keplerian component in the modelling, the power spectrum of the \drv~residuals contained significant excess power at $10.0$~d, i.e.\ the planet's period - a telltale sign that the periodic \drv~variations could not be explained by rotational activity alone.  When not including a Keplerian component in cases (ii) and (iii), visual inspection of the periodogram of residuals did not by itself point clearly to the presence of a planet with period $25.05$~d (the same as the stellar rotation period); however, the reality of the Keplerian components could be inferred by the significant improvement in the likelihoods of the models (approximated e.g.\ using Bayesian Information Criteria; see Section \ref{sec:tests-howard}) that included a planet}.

Thus we have demonstrated that our GP framework is able to disentangle activity signals from Keplerian signals, even when the Keplerian signal has a period identical to that of the activity signal, and an amplitude much smaller than the activity signal (close to the noise floor, in fact). Although the example we presented featured imperfect removal of the activity signal from the \drv~data (since the residuals contained a signature of one of the activity signal's harmonics), the ancillary time series served the role of constraining very tightly the activity component in the \drv~time series, allowing the Keplerian signal to be modelled accurately. Presumably, a more physically-realistic incarnation of our GP framework (see Section \ref{sec:discuss}) will allow activity signals to be even more tightly constrained by the ancillary time series. We also note that the harmonic of the activity signal was not present in the \drv~residuals when modelling the signal arising from the same configuration but with more data, as in Fig.\ \ref{fig:SOAP_002b}, where nearly four years' worth of data were included in the modelling. Nevertheless, the presence of a correlated, periodic artifact of the activity signal in the RV residuals in this test underscores the importance of accurate activity modelling, and the dangers of hastening to a planetary interpretation of signals in residuals. It would be worth checking, in future work, whether a Bayesian model comparison test would clearly penalise and disfavour a model that included a second Keplerian component to explain this signal.

We conclude the discussion of our \mbox{SOAP~2.0} tests by noting the following shortcomings of these tests, along with possibilities for future extensions.
\begin{enumerate}
\item Though the covariance kernel we used can naturally accommodate quasi-periodic signals (see Fig.\ \ref{fig:QP_example}), our simulations did not include any spot evolution; the \mbox{SOAP 2.0} code does not directly facilitate such evolution. In principle, though, activity signals arising from different spot configurations could be combined, perhaps with large gaps in between the different signals, to simulate evolving active regions. \textcolor{black}{Long-term magnetic cycles could be simulated in a similar fashion.}
\item Non-informative (uniform and Jeffreys) priors were placed on all model parameters/hyper-parameters. Improved fitting could be expected if we were to use priors informed by a better understanding of e.g.\ the hyper-parameters of our quasi-periodic covariance kernel, or more physically-motivated priors for Keplerian signals.
\item We did not include any plages in the active regions simulated by \mbox{SOAP 2.0}. At the time of writing, the \mbox{SOAP 2.0} code's output appeared to include a non-trivial amount of high-frequency noise when including plages in the simulations. Preliminary tests with a smoothing filter applied to the \mbox{SOAP 2.0} output suggest that even when plages are included, the activity components in \drv~time series can still be modelled accurately, although the large residuals in the ancillary time series in these tests underscore the simplicity of the physical model used in our framework. Complications arising from the use of smoothing filters notwithstanding, these tests suggest the need for a better understanding of the shortcomings of the physical model (ultimately to be replaced, hopefully, by a more sophisticated one) currently incorporated into our GP framework.
\end{enumerate}

\begin{table*}
  \centering
  \caption{MAP parameter estimates for selected Keplerian orbital elements in the planet injection tests, along with $\pm\sigma$ posterior uncertainties. Parameter inference was performed using the MultiNest nested-sampling algorithm, with non-informative priors placed on all GP mean function (Keplerian) parameters. }
    \begin{tabular}{cccccccc}
\hline
          & \multicolumn{2}{c}{$\Delta$~RV semi-amplitude $K$ [m/s]} & \multicolumn{2}{c}{Period $P_K$ [days]} & \multicolumn{2}{c}{Eccentricity $e$} 
          \\
          \cline{2-7}
          & \multicolumn{1}{c}{Model} & \multicolumn{1}{c}{Truth} & \multicolumn{1}{c}{Model} & \multicolumn{1}{c}{Truth} & \multicolumn{1}{c}{Model} & \multicolumn{1}{c}{Truth} \\
          \hline
    Case (i) & \multicolumn{1}{c}{$1.400\pm0.045$} & \multicolumn{1}{c}{$1.40$} & \multicolumn{1}{c}{$9.9941 \pm 0.0086$} & \multicolumn{1}{c}{$10.0$} & \multicolumn{1}{c}{$0.118\pm0.016$} & \multicolumn{1}{c}{$0.10$} \\
    Case (ii) & \multicolumn{1}{c}{$1.372 \pm 0.070$} & \multicolumn{1}{c}{$1.40$} & \multicolumn{1}{c}{$25.09\pm0.26$} & \multicolumn{1}{c}{$25.05$} & \multicolumn{1}{c}{$0.1050 \pm 0.0081$} & \multicolumn{1}{c}{$0.10$} \\
    Case (iii) & \multicolumn{1}{c}{$0.334\pm0.047$} & \multicolumn{1}{c}{$0.28$} & \multicolumn{1}{c}{$25.10\pm0.89$} & \multicolumn{1}{c}{$25.05$} & \multicolumn{1}{c}{$0.1051\pm0.0080$} & \multicolumn{1}{c}{$0.10$} \\
          \hline
    \end{tabular}%
  \label{tab:planet_fit}%
\end{table*}%

\subsection{The planet around Gl~15~A}\label{sec:tests-howard}
\citet{howard2014} -- hereafter H14 -- reported the discovery of low-mass planet orbiting Gl~15~A, part of the Gl~15 binary system, based on RVs from the Eta-Earth survey using HIRES. They characterise Gl~15~Ab as a planet with minimum mass $M\sin i = 5.35 \pm 0.75M_\oplus$, based on an RV semi-amplitude of $K=2.94\pm0.28$~\mps. The planet has an orbital period $P_K=11.4433\pm0.0016$~d, and an orbit that is consistent with circular. The detection and characterisation of Gl~15~Ab was based on data with a 15-year baseline (1997 January through 2011 December).

Gl~15~A itself is a modestly-active, M2~V dwarf. Over their 15-year baseline, H14 detected a $9\pm2.5$ year activity cycle with a semi-amplitude of $\sim0.05$, in the dimensionless units of $S_\textrm{HK}$ (derived from \lrhk), which represents a $\sim10\%$ fractional change; this variation may be a magnetic cycle analogous to the solar cycle. On shorter time-scales, they detected a strong periodicity in $S_\textrm{HK}$ measurements with period near $44$~d, which they interpreted as rotationally-modulated activity. The fact that they also detected these $\sim44$~d modulations in optical photometry and RV time series supported this interpretation. On the other hand, the $11.44$~d RV signal was not detected in photometry or $S_\textrm{HK}$, strengthening the planetary interpretation of the $11.44$~d signal.

Since the $11.44$~d period of Gl~15~Ab could easily be mistaken for the fourth harmonic of the $\sim44$~d stellar rotation period, H14's publicly-available dataset provides an interesting test case for our GP model: an obvious (though perhaps unfounded) concern in this case would be that the GP activity model would simply absorb the planetary signal.

 Accordingly, we modelled the 15 years of data published by H14, comprising $117$ RV and $S_\textrm{HK}$ measurements, 59 of which were taken in the 2011 season (BJD $>$\num[group-separator={,}]{2455500}). As for our earlier tests, we used a quasi-periodic covariance kernel for the latent process driving activity; we also included a linear RV trend in both our single-planet and planet-free models, to account for the $\sim2~$\mps~yr$^{-1}$ acceleration of Gl~15~A by its companion Gl~15~B.

When we did not include a Keplerian component in the mean function for the \drv~time series, the Lomb-Scargle periodogram of the residuals contained multiple significant peaks, including one at around $11.4$~days, indicating that the putative planetary signal was not absorbed by the activity model: since this periodic modulation was not present in the $S_\textrm{HK}$ time series, our model did not allow it in the \drv~time series. On the other hand, when including a Keplerian component with circular orbit in the \drv~mean function, the quality of the fit improved significantly, and a Lomb-Scargle periodogram of the residuals indicated that no significant periodicities remained. Using Bayesian Information Criteria (BIC) as a first-order approximation to Bayes factors \citep{berger1996,raftery99}, our single planet model was strongly preferred with $\Delta\textrm{BIC}=27$; H14 arrived at the same conclusion, with $\Delta\textrm{BIC}=24$.

\begin{figure*}
\begin{center}
\includegraphics[width=\textwidth]{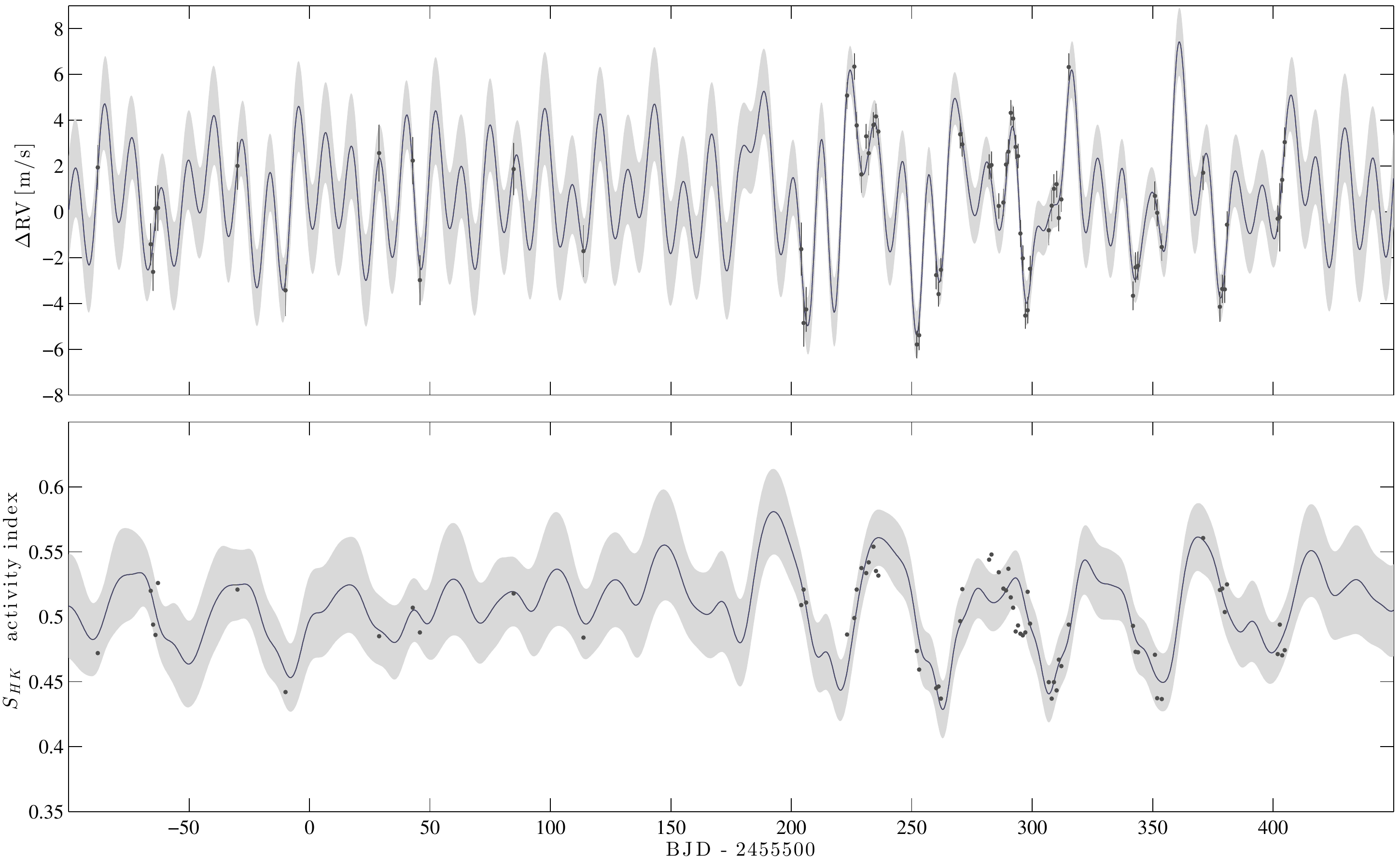}
\caption[]{GP model MAP fit to the publicly-available Gl~15~A data from H14. The fit was performed to the full 15 years of data, comprising $117$ observations; since most of the observations were made during the 2011 season (BJD $>$\num[group-separator={,}]{2455500}), however, only data from this season are plotted here. The dots indicate H14's data (with estimated errors, where applicable); the solid lines are model posterior means, and the shaded regions denote $\pm\sigma$ posterior uncertainty.}
\label{fig:howard_A}
\end{center}
\end{figure*}

\begin{figure*}
\begin{center}
\includegraphics[width=\textwidth]{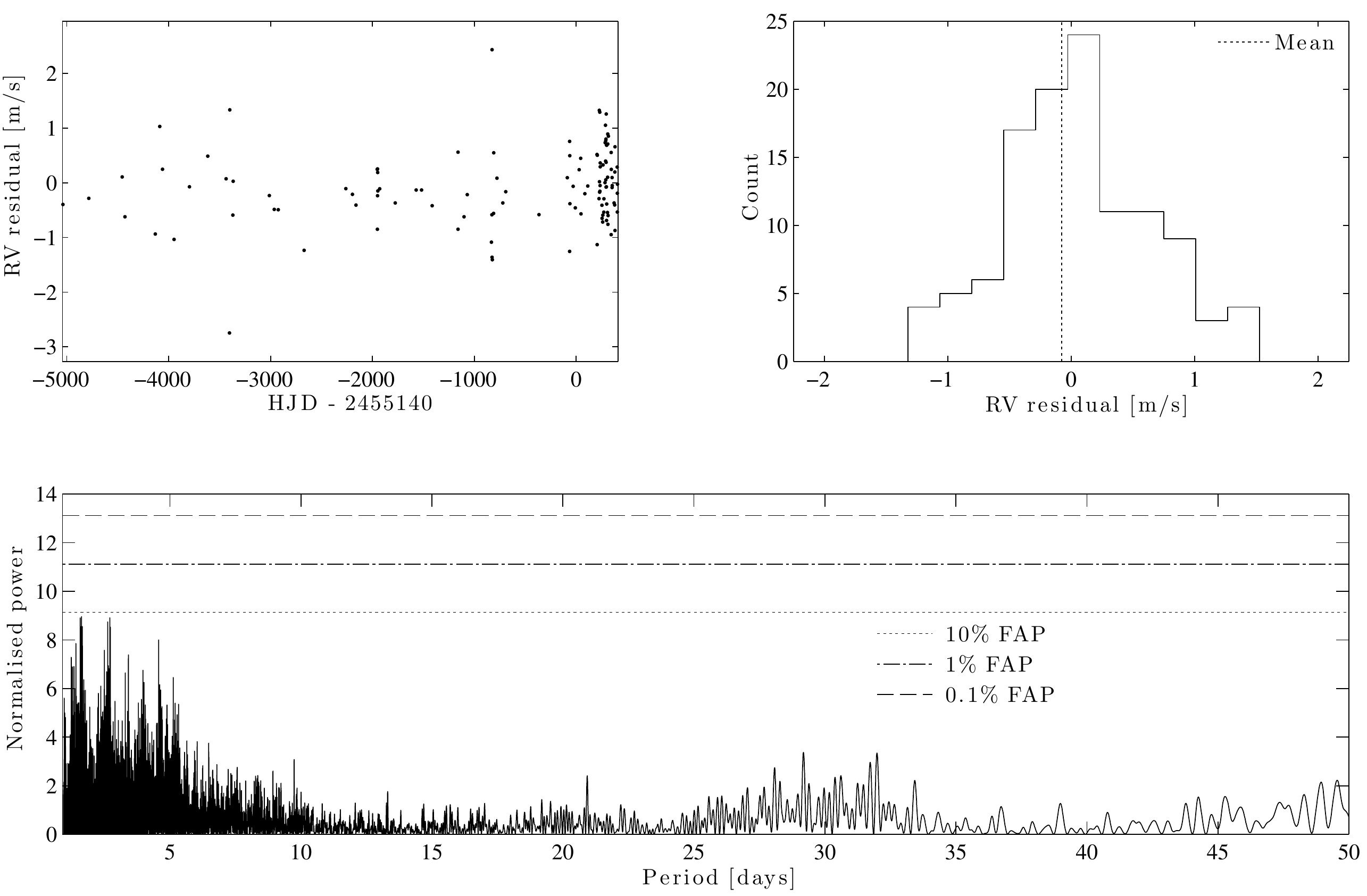}
\caption[]{Diagnostics of the \drv~residuals for Gl~15~A, after subtracting our MAP GP model (stellar activity, plus binary orbit and one planet). The top left panel contains the residuals plotted as a function of time, spanning some 15 years; the top right panel contains a histogram of the residuals (rms: $0.69$~\mps); and the bottom panel shows the normalised Lomb-Scargle periodogram of the residuals, along with false alarm probability thresholds.}
\label{fig:howard_B}
\end{center}
\end{figure*}

Our GP model, one-planet fit to H14's data is presented in Fig.\ \ref{fig:howard_A}, while detailed diagnostics of the \drv~residuals (rms: $0.69$~\mps) are presented in Fig.\ \ref{fig:howard_B}. Our MAP estimate for the \drv~semi-amplitude of the planetary signal was $K=2.22\pm0.36$~\mps, and our estimate for its period $P_K=11.4431\pm0.0010$~d. \textcolor{black}{H14 obtained $K=2.94\pm0.28$~\mps, and $P_K=11.4433\pm0.0016$~d}; thus our approach to modelling H14's dataset led us to detect a planet with period and amplitude consistent with those derived by H14, to within our respective $95\%$ ($2\sigma$) credible intervals for these parameters. The MAP estimate of the period hyper-parameter in our quasi-periodic covariance kernel was $P=43.8\pm0.9$~d, which is consistent with the $\sim44$~d stellar rotation period derived by H14.

We did not consider any multiple-planet models. We defer such analyses of this and other datasets to future work, in which we aim to leverage a better understanding of appropriate priors over our GP model's hyper-parameters to perform more rigorous Bayesian model comparison.

\subsection{The $\boldsymbol{\alpha}$~Cen~B dataset}\label{sec:tests-alphaCen}
The announcement by \citet{dumusque2012nature} of the detection of a planet around the modestly-active, \mbox{K1 V} star \aCB~caused a major stir in the exoplanet community: if verified, the claimed planet \aCB b would be the closest (distance: $1.34$~pc) exoplanet to Earth ever discovered, and the lowest-minimum-mass planet detected around a Solar-type star. In particular, the claimed planet has an orbital period of $3.24$~d, and a minimum mass of $1.13\pm0.09$~$M_\oplus$.

\citeauthor{dumusque2012nature} (hereafter D12) obtained $459$ HARPS RV datapoints, along with ancillary FWHM, \bis\ and \lrhk\ time series, over a period of 4 years. The D12 RVs are dominated by a long-term linear trend, which is due to the orbit of \aCB\ around the centre of mass of the $\alpha\,{\rm Cen}$ binary system. Once this trend is subtracted, a gradual rise and fall over the 4-year span of the observations is evident, as well as variability on shorter time-scales. D12 interpreted the gradual rise and fall as a signature of the star's activity cycle, and the quasi-periodic variations on time-scales of a few weeks as caused by the rotational modulation of star spots. Lastly, there are short-term variations evident in all time series. These variations are correlated in time, with a time-scale of a few days, as well as with each other (across all three time series), albeit in a somewhat complex way. It is unclear whether these variations might be due to activity, observational noise, or a combination of both. Of course, in the case of the RV data in isolation, a part of this variability could also be explained by a planetary companion with a relatively short orbital period.

In any event, D12 used a variety of mathematical transformations to try to filter out many sources of RV variance -- including starspots, photospheric granulation, and binary motion due to the presence of companion star \aCA\ -- to isolate the putative planet's RV signal. The amplitude of the binary-motion signal they removed was on the order of hundreds of \mps; the combined amplitude of the nuisance signals ascribed to stellar activity they removed was on the order of $10$~\mps; and the final, claimed planetary signal they isolated had a semi-amplitude of about $\sim0.5$~\mps. For comparison, the long-term precision of HARPS is $0.8$~\mps \citep{mayor2003}.

However, the detection was a contentious one, with a number of authors calling into question the planet's existence \citep[see e.g.][]{hatzes2012,hatzes2013}. Some of the possible reservations about the approach used by \citeauthor{dumusque2012nature} include the following.
\begin{enumerate}
\item Short-term stellar rotational activity was dealt with by fitting sinusoidal waves at the rotational period of the star, and a variable number of harmonics. The constraint that all activity-induced variations should be strictly sinusoidal is very hard to motivate physically \citep{lanza2001,brinkworth2005}, and subtracting multiple sinusoids from the \drv~time series opens the possibility of inadvertently introducing periodic signals (or harmonics thereof) that weren't present in the original data. 
\item Available activity-sensitive time series (BIS and $\log{R'_\textrm{HK}}$) were not jointly modelled with the $\Delta$RV time series -- even though it is in the \lrhk\ time series that the rotational signal was most {prominent}. Instead, a low-pass smoothing filter was applied to the $\log{R'_\textrm{HK}}$ time series; it was then simply noted that the smoothed time series looked similar to the $\Delta$RV time series, and the shape of this smoothed time series was then used to fit the radial velocity variations in order to mitigate the effects of long-term activity.
\item Each of the four observing seasons was fitted independently of the others, thus ignoring any information about possible long-term trends/correlations in the time series, and resulting in a model with almost four times as many degrees of freedom as might've been the case if fitting were performed for all seasons simultaneously.
\item The model used to fit the $\Delta$RV time series alone, \emph{without} a planet, contained $23$~free parameters, and the combined fit for activity and a planet contained at least $26$~free parameters. The question of possible over-fitting was not addressed in either case. For comparison, our baseline/planet-free GP model -- discussed below -- models multiple time series simultaneously, yet contains significantly fewer free parameters than D12's model.\footnote{Of course, a more meaningful application of Occam's razor would require marginalisation over the full volumes of respective model parameter spaces.}
\item No rigorous model comparison was performed to quantify the evidence for the existence of the planet, i.e.\ the extent to which a planetary model is (or isn't) preferred.
\end{enumerate}

Given that our GP framework for modelling activity would allows us, in principle, to avoid all of these limitations, we decided to apply it to the publicly-available D12 dataset.

As in all previous tests, we used a quasi-periodic covariance kernel for our latent process. We modelled all $459$ datapoints in each of the \drv, \lrhk and \bis~time series  jointly using this single latent process and its derivative.\footnote{The FWHM data are noisier than, but very tightly correlated with, \lrhk, and thus are {unlikely} to contain useful extra information. Both can be seen as proxies for the integrated spot coverage of the visible hemisphere. On the other hand, the \bis~also depends on the velocity of the stellar surface at the location of the spots (or active regions).} \textcolor{black}{The \drv~time series was corrected for binary motion, with a residual DC offset removed for each season (in general we obtained identical results whether we removed long-term drifts prior to the GP modelling, or included extra terms in the GP mean functions to take these into account; as a computational convenience to reduce the dimensionality of our hyperparameter space, then, we opted for the former approach).} Error estimates were published by D12 for the \drv~and \lrhk~time series; these estimates were included in our modelling. Though not directly provided, BIS errors were estimated (following D12's prescription) using provided photon-noise estimates. Finally, as before, non-informative priors were placed on all model (hyper)parameters.

\begin{figure*}
\begin{center}
\includegraphics[width=\textwidth]{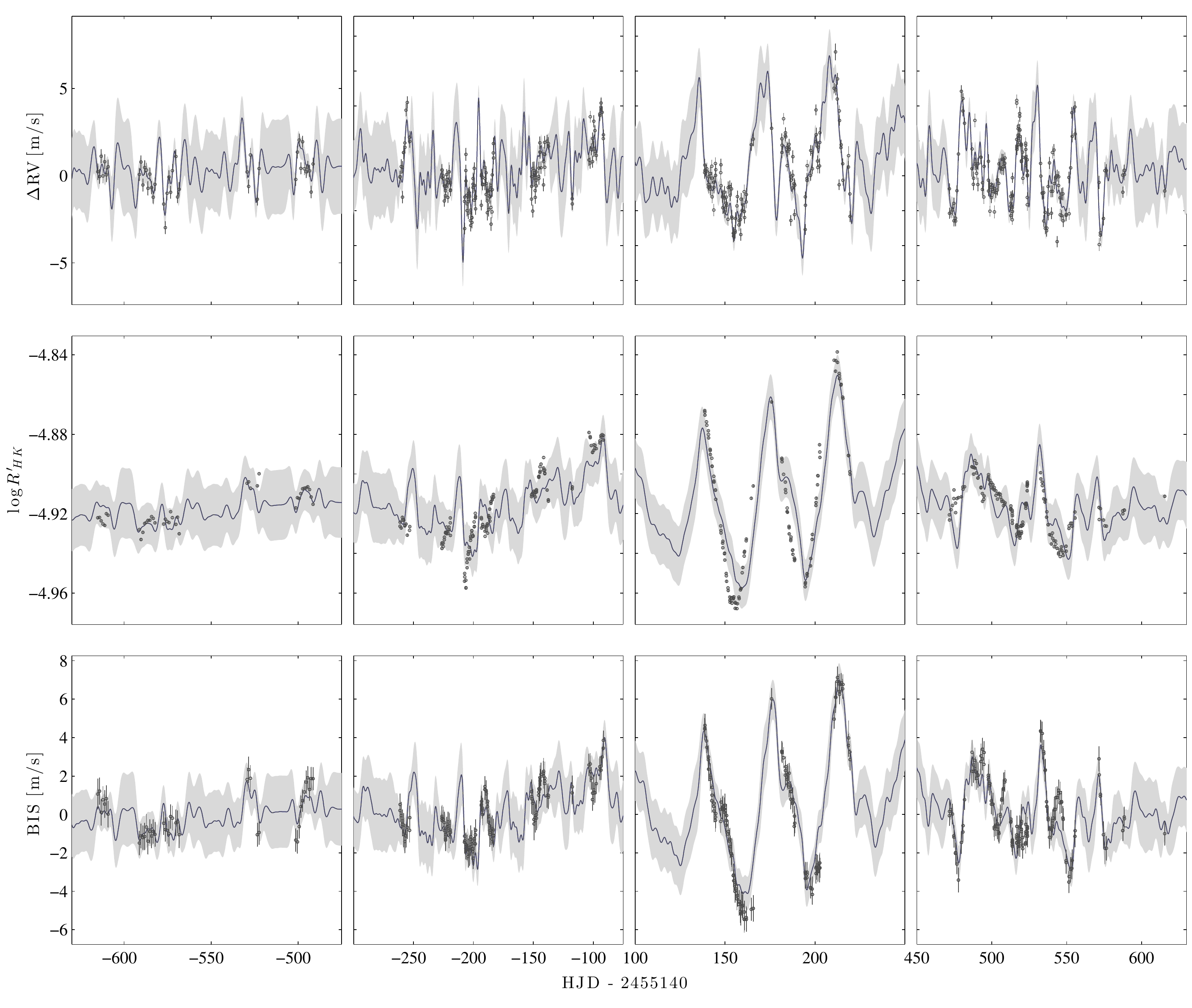}
\caption[]{GP model MAP fit to the publicly-available \aCB~data from D12. All four seasons, comprising some $459$ observations, were fit simultaneously, i.e.\ using a single set of (hyper)parameters. The dots indicate D12's data (with estimated errors, where applicable); the solid lines are model posterior means, and the shaded regions denote $\pm\sigma$ posterior uncertainty.}
\label{fig:Alpha_A}
\end{center}
\end{figure*}

\begin{figure*}
\begin{center}
\includegraphics[width=\textwidth]{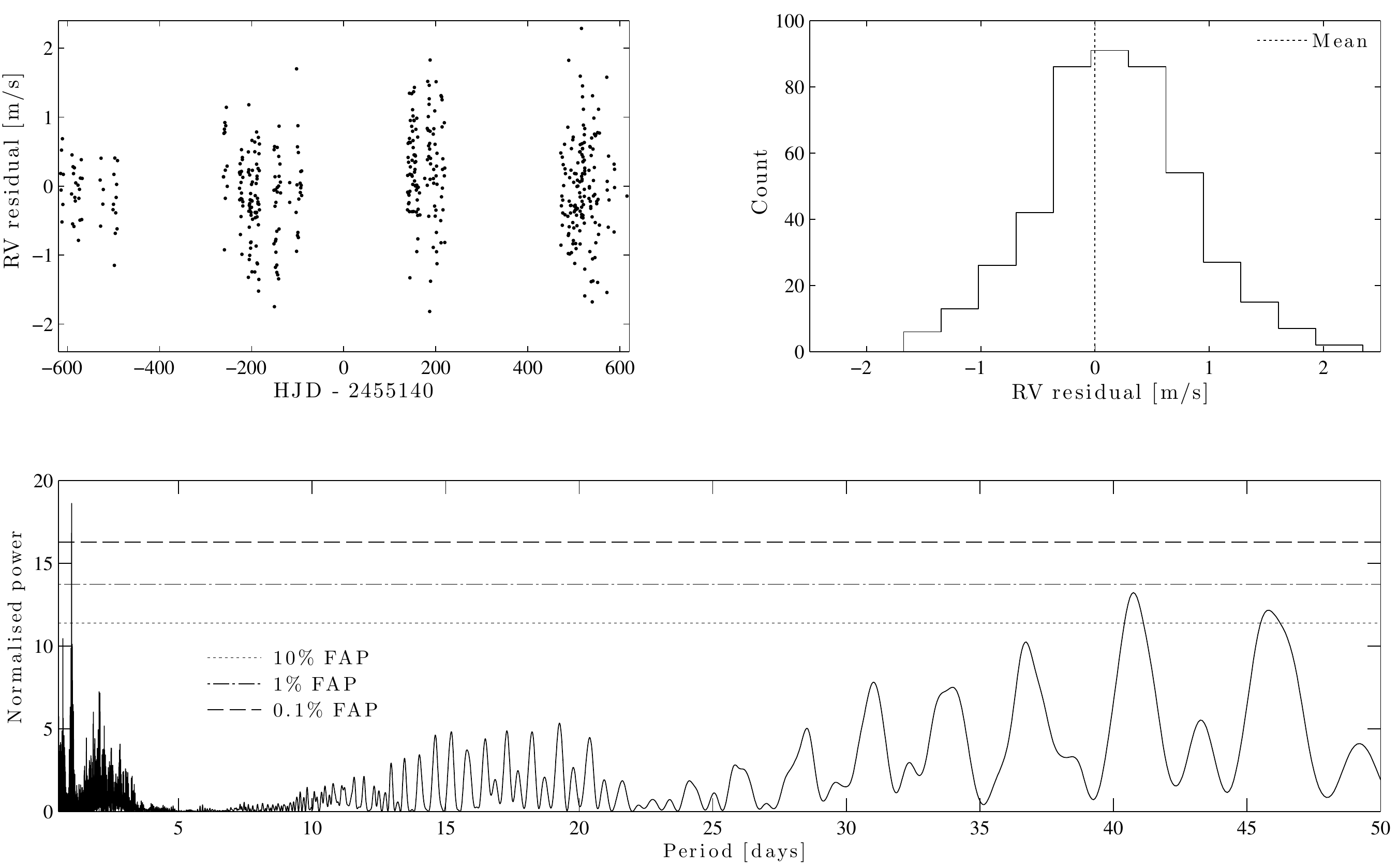}
\caption[]{Diagnostics of the \drv~residuals for \aCB, after subtracting our MAP GP model (stellar activity plus binary orbit). The top left panel contains the residuals plotted as a function of time the top right panel contains a histogram of the residuals (rms: $0.70$~\mps); and the bottom panel shows the normalised Lomb-Scargle periodogram of the residuals, along with false alarm probability thresholds.}
\label{fig:Alpha_B}
\end{center}
\end{figure*}

Our baseline GP model did not incorporate any Keplerian component in the mean function for the RVs. In total, the combined model for all three time series contained 14~free (hyper)parameters -- in particular, it should be noted that all three time series across all four observing seasons were fitted using this same model. This stands in contrast to the approach used by D12, where the activity signal in each of the four seasons was fitted independently, since the model they used required that stellar features did not evolve significantly during the time considered, whereas spots and plages and spots are known to evolve on a time-scale of a few rotations. The  MAP fit using this model is presented in Fig.\ \ref{fig:Alpha_A}, while detailed diagnostics of the \drv~residuals are presented in Fig.\ \ref{fig:Alpha_B}. 

The \drv~residuals appear to be normally-distributed, with rms $0.70$~\mps. \textcolor{black}{The only significant features in the power spectrum of the \drv~residuals, presented in Fig.\ \ref{fig:Alpha_B}, is a complex of peaks tightly clustered around $1.00$~d ($<0.1\%$ false alarm probability), and a broad complex of peaks clustered near the $\sim37.8$~d stellar rotation period ($>1\%$ false alarm probability). The lack of other peaks in the residuals suggests that the \drv~variations can be explained as arising from activity alone.}

\textcolor{black}{The presence of residual power near the stellar rotation period indicates that our framework does not do a perfect job of describing the rotationally-modulated activity signals -- at least, not when fitting all four seasons with a single model. This might be indicative of either differential rotation, as suggested by D12, or of spot pattern evolution; a more detailed investigation of this possibility is deferred to a future study.}

As noted in Section \ref{sec:tests-SOAP}, the peaks around $1.00$~d arise from the window function in the \aCB~dataset. In particular, there are peaks at $f_1=0.97$~d$^{-1}$ and $f_2=1.03$~d$^{-1}$; these peaks appear in the power spectra of both the model and the observations, and may be interpreted as the \textcolor{black}{first-order daily aliases} of the $1/37.8$~d$^{-1}$ rotational frequency. \textcolor{black}{A third peak at $f_3=1.00$~d$^{-1}$ appears only in the \drv~residuals, but not in the model or the observations: this may be interpreted as a beat frequency arising during the computation of residuals, on account of interference between the other two closely-spaced frequencies, $f_1$ and $f_2$.}\footnote{\textcolor{black}{To understand the beat phenomenon, it is useful to recall the trigonometric identity: $\sin (2\pi {f_1}t) \pm \sin (2\pi {f_2}t) = 2\sin \left( {2\pi \frac{{f1 \pm f2}}{2}t} \right)\left( {2\pi \frac{{f1 \mp f2}}{2}t} \right)$.}}

When we did include a Keplerian component in the mean function for the RVs, we were unable to extract any significant (non-activity) signal. Even when forcing the Keplerian signal to have a period of close to $3.24$~d, the MAP solution favoured a Keplerian signal with an arbitrarily small (within the bounds of a Jeffreys prior) amplitude. Conversely, when forcing the Keplerian signal to have an amplitude of about $\sim0.5$~\mps, the MAP solution favoured either a very short ($P_K\ll1$~d) or very long ($P_K\gg1$~year period), so that this signal essentially became a DC offset to the \drv~time series. Furthermore, our non-detection of the planetary signal was also robust against different choices of covariance kernel functions (e.g.\ squared exponential, rational quadratic). Nor did allowing for somewhat tighter or looser links between the RVs and the activity-sensitive \lrhk\ and \bis\ time series, by modifying the priors on the three $\sigma_{l_i}$ terms in our model, allow us to detect a $3.24$~d Keplerian signal in the data. 

In summary, using the activity-sensitive \lrhk\ and \bis\ time series to constrain the activity component of the RV time series suggests that all of the significant variation in the RV time series can be explained as being induced by stellar activity, without requiring any additional Keplerian component. 

We do not claim that this finding demonstrates that the claimed planet \aCB b definitely does not exist; however, the difficulty we have in recovering the signal as something unrelated to activity underscores just how important it is to model activity carefully and robustly, if we are to detect and characterise planets at the sub-\mps-level.

\begin{figure*}
\begin{center}
\includegraphics[width=\textwidth]{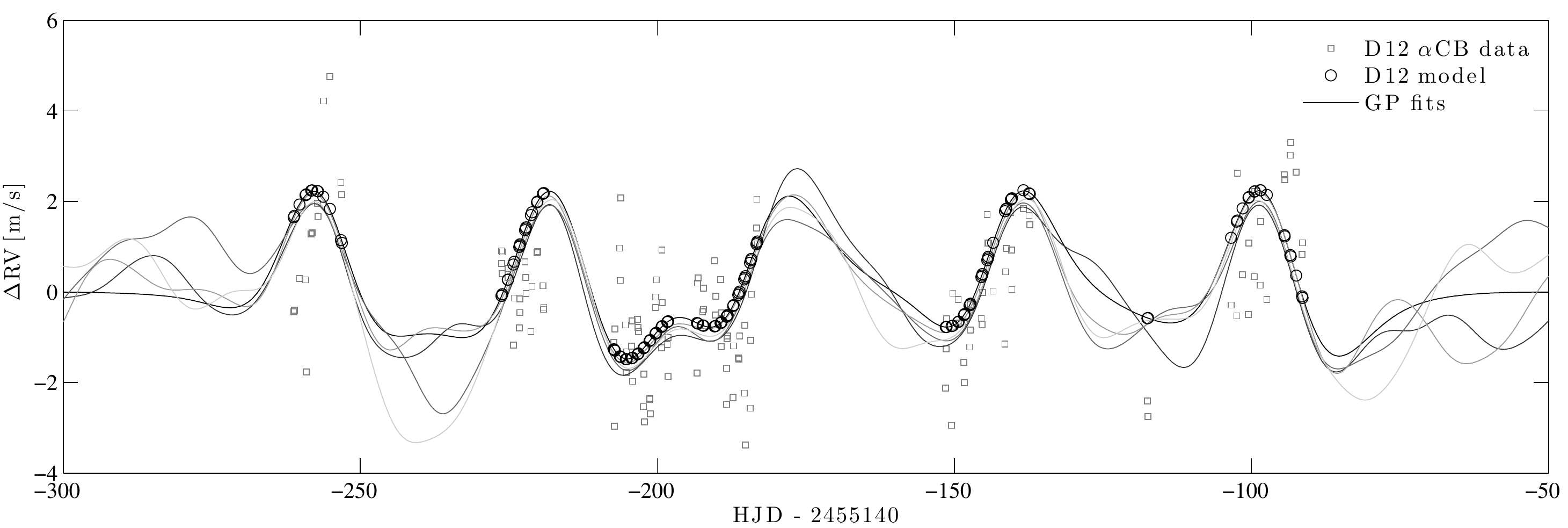}
\caption[]{\textcolor{black}{Demonstration that the predictions of a GP model can subsume those of D12's model as a special case. The grey squares correspond to the publicly-available \aCB~data from D12, after removing binary and long-term magnetic cycle contributions, while the black circles are predictions from D12's published activity model, for the same times as the real observations. Draws from a GP model (a few examples of which are plotted) fitted to the noisy data can be found such that they essentially interpolate D12's model predictions; for visual clarity, only one observing season is shown here, but the interpolation holds true for the other seasons as well. Indeed, it is straightforward (though not necessarily useful or desirable, and ignoring considerations of over-fitting) to find hyper-parameters that will force a GP model to agree exactly with D12's activity model, for arbitrarily-many observing times.}}
\label{fig:D12_model}
\end{center}
\end{figure*}

In the near future, we aim to make a more rigorous and comprehensive study of D12's \aCB~dataset. We would like to investigate in which situations our planet-free models lead to a signal with a $3.24$~d period in the RV residuals. This should happen at least in some cases, since our GP framework should be able to reproduce the model of \citeauthor{dumusque2012nature} as a special case -- even if only with a non-optimal model \textcolor{black}{(see Fig.\ \ref{fig:D12_model})}, and/or when loosening significantly the links between the ancillary, activity-sensitive time series and the RVs. Concurrently, we would like to undertake detailed studies of planetary detection limits using both our GP framework and a model akin to that used by \citeauthor{dumusque2012nature}, for synthetic data similar to the \aCB\ dataset.  \textcolor{black}{To draw definitive conclusions about whether the data really suggest the presence of a planetary signal, we would like to perform rigorous Bayesian model comparison, i.e.\ of planet vs.\ no-planet models. Finally, planetary considerations aside, we would like to investigate whether the inability of our model to remove all of the rotationally-modulated activity in the \drv~time series might be indicative of either differential rotation, or spot pattern evolution; one approach might be to use our GP framework to model each of the four seasons of data independently, and to study the changes (if any) in characteristic periodicity for each season.} These analyses will form the focus of a separate paper.

In the longer term, more measurements with denser time sampling -- coupled with a better understanding of the noise characteristics of the RV data -- could also lead to more definitive conclusions about the existence of \aCB b.
\section{Discussion and conclusions}\label{sec:discuss}
Stellar activity can induce RV variations that can drown out or even mimic planetary signals, and it is notoriously difficult to model and thus mitigate the effects of these activity-induced nuisance signals. This is expected to be a major obstacle to using next-generation spectrographs to detect lower and lower mass planets, planets with longer periods, and planets around more active stars. We have presented here a new framework for modelling RV time series jointly with one or more ancillary, activity-sensitive time series (including photometry, line widths, chromospheric activity indices, line asymmetries, etc.), with a view to better constraining activity signals in RVs, and thus better detecting and characterising possible planets. 

Our framework treats the underlying stochastic process giving rise to activity signals in all available observables (RVs and ancillary time series) as being described by a GP, with suitably-chosen covariance function. We then use physically-motivated and empirical models to link this GP to the observables; with the addition of noise and deterministic components (e.g.\ dynamical effects for the RVs), all observables can be modelled jointly as GPs, with the ancillary time series thus serving to constrain the activity component of the RVs.

We demonstrated the performance of our framework using both synthetic (Sections \ref{sec:tests-GP} and \ref{sec:tests-SOAP}) and real (Sections \ref{sec:tests-howard} and \ref{sec:tests-alphaCen}) data. We started by noting that our framework can model all available time series jointly, and exactly, in the simplest case where the physical/empirical relationships between all time series holds exactly. Next, using more realistic data simulated using the \mbox{SOAP~2.0} tool (including noise and realistic time sampling), we showed that our framework does a good job of constraining activity signals in RV data, provided we allow for additive white-noise terms in each time series we model; these additive noise terms may be interpreted as accounting for the extent to which our assumed relationships between the time series does not hold exactly, observational noise notwithstanding. We also showed that the framework can be used to disentangle activity and planetary signals -- even when the planetary signal is weaker than and has a period identical to the activity signal -- thus serving its intended purpose. Moving to real data, we applied our framework to the \mbox{Gl~15~A} system; we were able to disentangle activity and planetary components in a HIRES RV dataset, and obtained a fitted planetary model which was consistent with one published by \citet{howard2014}. Finally, we turned our attention to the much-discussed \aCB~dataset: we showed that, contrary to the analysis of \citet{dumusque2012nature}, we were able to attribute the observed radial velocity variations to stellar activity, without requiring a planet.

Although our framework hinges, at least partly, upon a number of approximations and empirical relationships, its performance appears promising, and it offers a number of advantages over existing approaches to mitigating activity in RV datasets. It is flexible; though algebraically non-trivial, it is conceptually simple, and makes minimal assumptions about the properties of the underlying processes inducing activity signals in observables; it facilitates smooth interpolation between observables, as well as extrapolation to future times (prediction); and lastly, the entire framework is accommodated very naturally within the broader framework of Bayesian inference. As such, we hope that our new framework will form a useful addition to the toolbox of those confronted with mitigating stellar activity signals in RV datasets.

An investigation of ways in which our framework could be improved would represent a natural extension to this work. Some possibilities worth studying include the following.
\begin{enumerate}
\item Developing a more realistic relationship between RV and ancillary time series, especially the \bis\ time series -- perhaps based on theoretical considerations, or perhaps through empirical means, e.g.\ by studying simulated data. Preliminary tests suggest the inclusion of a curvature term (second derivative of the latent process) could be helpful for more accurately coupling the RV and BIS time series; though it would add significant algebraic complexity, it would only add one free parameter to the overall model.
\item Performing more comprehensive studies of planet detection rates and false-positive rates under our framework.
\item Extending the \mbox{SOAP 2.0} tests to include plages and spot evolution.
\item Investigation of a more physically-motivated GP covariance function, e.g.\ a function from the Mat\'ern class of covariance functions, which are differentiable only a finite number of times, and have the squared exponential function as a smooth limiting case \citep{rasmussen2006}.
\item Investigation of more physically-motivated priors for all model (hyper)parameters, with a view to computing Bayes factors and performing rigorous model comparison. \textcolor{black}{In particular, this will require marginalisation over GP hyper-parameters, rather than fixing them at their MAP values.}
\item Using raw, high-resolution spectra to construct a quantity that is a more sensitive proxy to stellar activity than ones currently in widespread use (BIS, \lrhk, etc.).  An artificial neural network might be useful for machine-learning such a quantity from a large dataset.
\item \textcolor{black}{Investigation of including a GP component to model correlated instrumental noise, rather than forcing such noise to be artificially absorbed by an additive white-noise component, as is the case in our current framework.}
\end{enumerate}
In tandem with the above, as discussed in Section \ref{sec:tests-alphaCen}, we aim to present in the near future a far more detailed and rigorous analysis of the much-discussed \aCB\ dataset. We would like to investigate in which situations our fitted planet-free models lead to a signal with a $3.24$~d period in the RV residuals (this should happen at least in some cases, since our framework should be able to reproduce the model of \citeauthor{dumusque2012nature} as a special case); we would like to undertake detailed studies of planetary detection limits using both our GP framework and a model akin to that used by \citeauthor{dumusque2012nature}, for synthetic data similar to the \aCB\ dataset; and we would like to perform Bayesian model comparison (planet vs.\ no-planet models) in order to draw definitive conclusions about whether the data really suggest the presence of a planetary signal. These analyses will form the focus of a separate paper.

\section*{Acknowledgments}
The authors wish to thank Xavier Dumusque, Andrew Collier Cameron, Rapha\"elle Haywood and Heather Cegla for useful discussions. V.~R.\ acknowledges with gratitude Merton College, the Rhodes Trust, and the National Research Foundation of South Africa for providing financial support for this work.

\bibliography{biblio}
\bibliographystyle{mn2e}
\appendix
\section{Covariance between observations and derivative observations}
\textcolor{black}{Note that some of the equations given below already appear, albeit without further explanation, in the main body of the paper. The aim of this appendix is to provide some further details to allow these covariance kernels and their derivatives to be implemented computationally.}

Suppose we have a Gaussian process $G$ characterised by a covariance function $\gamma$. Then the covariance between an observation of $G$ at time $t_i$, and an observation of its derivative $\dot{G}$ at time $t_j$, is given by
\begin{equation}\label{eq:A1}
{{\gamma ^{(G,dG)}}({t_i},{t_j}) = \frac{\partial }{{\partial t}}{\left. {{\gamma ^{(G,G)}}(t,{t_j})} \right|_{t = {t_i}}},}
\end{equation}
{where ${{\gamma ^{(G,G)}}({t_i},{t_j})}$ is used to denote the covariance between (non-derivative) observations of $G$ at times $t_i$ and $t_j$. Similarly, the} covariance between two observations of {$\dot{G}$} at times {$t_i$ and $t_j$} is
\begin{equation}\label{eq:A2}
{{\gamma ^{(dG,dG)}}({t_i},{t_j}) = \frac{\partial }{{\partial t'}}\frac{\partial }{{\partial t}}{\left. {{{\left. {{\gamma ^{(G,G)}}(t,t')} \right|}_{t = {t_i}}}} \right|_{t' = {t_j}}}}.
\end{equation}

For convenience, we present below the relevant expressions for ${{\gamma ^{(G,dG)}}}$, ${{\gamma ^{(dG,dG)}}}$ etc.\ for the two covariance functions considered specifically in this paper. The above relations are, however, valid for any covariance function $\gamma$; \textcolor{black}{as such, the two examples below serve as an illustration of how the relevant expressions can be derived for any covariance function.}
\subsection{Squared-exponential covariance}\label{sec:SEappendix}
Using our `generalised squared-exponential' covariance function, the expression for the covariance between two observations of a process $G$ at times $t$ and $t'$ is:
\begin{equation}
\gamma _{se}^{(G,G)}(t,t') = \sum\limits_{i = 1}^N {{\beta _i^2}\exp \left[ { - \frac{{{{(t - t')}^2}}}{{2\lambda _i^2}}} \right]};
\end{equation}
the $\beta_i$'s control the relative amplitude of the $N\ge1$ components with evolutionary time-scales $\lambda_i$. For the case $N=1$, this reduces to the standard square-exponential covariance function.

Following equation \eqref{eq:A1}, the covariance between an observation of $G$ at time $t$ and an observation of $\dot{G}$ at time $t'$ is given by
\begin{equation}
\gamma _{se}^{(G,dG)}(t,t') =  - \sum\limits_{i = 1}^N {{\beta_i^2}\frac{{(t - t')}}{{\lambda _i^2}}\exp \left[ { - \frac{{{{(t - t')}^2}}}{{2\lambda _i^2}}} \right]},
\end{equation}
and, by equation \eqref{eq:A2}, the covariance between an observation of $G$ at time $t'$ and an observation of $\dot{G}$ at time $t$ is given by
\begin{equation}
\gamma _{se}^{(G,dG)}(t',t) =  - \gamma _{se}^{(G,dG)}(t,t').
\end{equation}
The covariance is symmetric, by definition, from which it follows that $\gamma _{se}^{(G,dG)}(t',t)=\gamma _{se}^{(dG,G)}(t,t')$. Finally, the covariance between two observations of $\dot{G}$ at times $t$ and $t'$ is given by
\begin{equation}
\gamma _{se}^{(dG,dG)}(t,t'){\text{ }} = \sum\limits_{i = 1}^N {{\beta_i^2}\left[ {\frac{1}{{\lambda _i^2}} - \frac{{{{(t - t')}^2}}}{{\lambda _i^4}}} \right]\,\exp \left[ { - \frac{{{{(t - t')}^2}}}{{2\lambda _i^2}}} \right]}.
\end{equation}
\subsection{Quasi-periodic covariance}\label{sec:QPappendix}
Using a quasi-periodic covariance function formed from a squared exponential kernel, the expression for the covariance between two observations of a process $G$ at times $t$ and $t'$ is:
\begin{equation}
 {\gamma _{qp}^{(G,G)}(t,t')} = \exp \left\{ - \frac{\sin^2 \left[\pi (t-t')/P\right]}{2 \lambda_{\rm p}^2} 
 - \frac{(t-t')^2}{2\lambda_{\rm e}^2} \right\},
\end{equation}
where $P$ and $\lambda_{\rm p}$ correspond to the period and length scale of the periodic component of the variations, and $\lambda_{\rm e}$ is an evolutionary time-scale. While $\lambda_{\rm e}$ has units of time, $\lambda_{\rm p}$ is dimensionless, as it is relative to $P$. 

Defining $\phi \equiv 2 \pi (t-t') / P$, we then obtain the expressions for the covariance between derivative and non-derivative observations:
\begin{equation}
 {\gamma _{qp}^{(G,dG)}(t,t') = g_{\rm qp}(t,t')} \, \left[ - \frac{\pi \sin \phi}{2 P
 \lambda_{\rm p}^2}- \frac{t-t'}{\lambda_{\rm
 e}^2} \right],
\end{equation}
and
\begin{equation}
 {\gamma _{qp}^{(G,dG)}(t',t) = - \gamma _{qp}^{(G,dG)}(t,t')};
\end{equation}
by symmetry of the covariance, $\gamma _{qp}^{(G,dG)}(t',t)=\gamma _{qp}^{(dG,G)}(t,t')$. Finally, the covariance between two observations of $\dot{G}$ at times $t$ and $t'$ is given by
\begin{displaymath}
 {\gamma _{qp}^{(dG,dG)}(t,t') = \gamma _{qp}^{(G,G)}(t,t')} ~ \times 
\end{displaymath}
\begin{displaymath}
 ~~~~~~~~ \left[ -\frac{\pi^2 \sin^2 \phi}{4 P^2 \lambda_{\rm p}^4} 
 + \frac{\pi^2 \cos \phi}{P^2 \lambda_{\rm p}^2}
 - \frac{{\phi} \sin \phi }{2 \lambda_{\rm p}^2 \lambda_{\rm e}^2} 
- \frac{(t-t')^2}{\lambda_{\rm e}^4} + \frac{1}{\lambda_{\rm e}^2} \right]. 
\end{displaymath}
\begin{equation}
\end{equation}

\bsp
\label{lastpage}
\end{document}